\pdfoutput=1
\documentclass[acmsmall,screen,nonacm]{acmart}

\AtBeginDocument{%
  }

\usepackage{amsmath}
\providecommand{\checkmark}{\ding{51}}
\usepackage{pifont}
\usepackage{array}       
\usepackage{multirow}    
\usepackage{makecell}    
\usepackage{colortbl}   
\usepackage{tikz}
\usetikzlibrary{arrows.meta, shapes.geometric, positioning, fit, backgrounds, decorations.pathreplacing}
\usepackage{adjustbox}
\usepackage{placeins}    
\hypersetup{hidelinks}

\begin{document}

\title{Toward Secure AI-Powered Penetration Testing Agents: Security Threats, Guardrails, and Architectural Perspectives}

\author{Rahul Dev T Y}
\orcid{0009-0008-3823-3770}
\affiliation{%
  \institution{National Institute of Technology Calicut}
  \department{Department of Computer Science and Engineering}
  \state{Kerala}
  \postcode{673601}
  \country{India}
}
\email{rahul\_m251216cs@nitc.ac.in}

\author{Hiran V. Nath}
\orcid{0000-0001-7881-4694}
\correspondingauthor
\affiliation{%
  \institution{National Institute of Technology Calicut}
  \department{Department of Computer Science and Engineering}
  \state{Kerala}
  \postcode{673601}
  \country{India}
}
\email{hiranvnath@nitc.ac.in}

\begin{abstract}
LLM-powered autonomous agents are transforming the penetration testing space with dynamic, multi-step offensive security workflows that require minimal supervision by humans. These agents leverage sophisticated reasoning abilities and external security tools to independently carry out reconnaissance, identify vulnerabilities, devise exploitation plans, and perform post-exploitation operations. But the ability to have persistent memory, to take actions in the real world, and to do long-horizon reasoning raises qualitatively different security concerns than traditional chat-based LLM systems. Existing guardrail mechanisms for conversational AI may not be sufficient to secure autonomous AI pentesting agents accordingly.

To address these issues, we carry out a comprehensive security analysis on autonomous AI-penetration testing agents. We systematically analyse representative agent architectures, characterise their trust boundaries and attack surfaces and propose a threat taxonomy that is aligned with the lifecycle and covers LLM lifecycle attacks, agent-architecture attacks and cross-cutting behavioural attacks. We analyse the limitations of existing guardrail mechanisms, identify key research gaps, and discuss future research directions for developing specialised, context-aware, and architecture-aware guardrails to secure next-generation AI-driven offensive security systems.
\end{abstract}



\maketitle

\section{Introduction}
\label{sec:intro}

The rapid growth of large scale software systems, cloud infrastructure and interconnected digital services has greatly increased the attack surface of modern computing environments. Penetration testing is a systematic, authorised, offensive security test that is critical to identifying exploitable vulnerabilities before an adversary does. But the scale, dynamism and technical breadth of modern systems has outstripped the ability of purely human-driven security assessments, creating a scalability problem and a persistent global shortage of skilled security practitioners.~\cite{walter2024red}

Recent advances in large language models (LLMs) have opened a new frontier
in offensive security automation. LLMs exhibit strong multi-step reasoning, self-correction ability, and the ability to produce structured attack plans from high-level goals. When combined with agent frameworks that allow access to real-world tools~\cite{fang2024llm, zhu2026teams} such as network scanners, exploit frameworks, shell interpreters, and web browsers, LLMs can act as autonomous cyber operators that monitor the state of target systems, plan multi-stage attack sequences, execute tools, interpret results, and adapt strategies based on environmental feedback.

This has led to the rise of a new class of systems: \emph{autonomous AI pentesting agents}. The state of this paradigm is exemplified by platforms like PentestGPT~\cite{deng2024pentestgpt}, AutoPentest~\cite{henke2025autopentest}, VulnBot~\cite{kong2025vulnbot}, PenHeal~\cite{huang2023penheal}, ARACNE~\cite{nieponice2025aracne}, AutoAttacker~\cite{xu2024autoattacker}, BreachSeek~\cite{alshehri2024breachseek}, Cochise~\cite{happe2026cochise}, PentestAgent~\cite{shen2025pentestagent}, HackSynth~\cite{muzsai2024hacksynth}, AutoPentester~\cite{ginige2025autopentester}, HackingBuddyGPT~\cite{happe2023getting}, and Incalmo~\cite{singer2025incalmo} which demonstrate that artificial intelligence can automate tasks ranging from reconnaissance and vulnerability discovery to exploit generation and post-exploitation operations.

The benefits of these systems are impressive. But they introduce a new, under-appreciated class of security hazard: the attack surface is the AI system itself. A pentesting agent that can be gamed – in its inputs, its memory, its tool interfaces, or its underlying model training – is not just an ineffective tool, but an actively dangerous one that can execute real world cyber actions against unintended targets under adversarial control.

A major and largely unrecognised challenge is that \textbf{existing LLM guardrail mechanisms were designed for conversational artificial intelligence (AI), not for autonomous offensive-security agents that can execute real-world cyber actions.} Today’s guardrails are based on the assumption of human supervision, single-turn interactions, trusted input channels, and text-only output. These assumptions are systematically violated by autonomous pentesting agents, which operate autonomously in multi-step campaigns, take input from adversarially-controlled target systems, and directly translate LLM output into shell commands and exploit tool invocations.

In this survey, we address this mismatch by proposing a \textbf{two-axis lifecycle-aligned taxonomy} based on the structure of the LLM security research literature. The taxonomy is arranged in two perpendicular axes.
The first axis---LLM Lifecycle Attacks (Cat. A--C)---includes vulnerabilities inherited from the base model pipeline: pre-training and corpus poisoning (Cat. A), fine-tuning and alignment attacks (Cat. B), and inference-time prompt attacks (Cat. C). These vulnerabilities exist in any system that utilises an LLM, regardless of agent topology. The second axis---\emph{Agent Architecture Attacks} (Cat.\,D--F)---concerns vulnerabilities that are not inherited but \emph{emergent}, i.e., they only emerge when the agentic system adds persistent memory or RAG (Cat.\,D), inter-agent communication (Cat.\,E), or external tool execution (Cat.\,F). This is reflected in a symmetric six-category guardrail taxonomy (G-A to G-F) For each guardrail category, this survey systematically characterises not only whether existing mechanisms provide coverage, but \emph{why} they are structurally insufficient in the pentesting agent context---a distinction not previously formalised in the literature.

\subsection{Contributions}

In brief, we present here the main contributions of this survey:
\begin{itemize}
\item We propose a unified analytical framework for security analysis of autonomous AI pentesting agents by systematically correlating architectural elements, trust boundaries, attack surfaces, failure modes, guardrail mechanisms and research challenges. The framework offers a systematic approach to examine how architectural design decisions impact the security stance of AI-based pentesting systems.

\item We present a comprehensive architecture analysis of representative AI pentesting agents. We include and analyse key components such as reasoning and planning modules, memory management, retrieval-augmented generation (RAG), tool integration, reflection mechanisms, and multi-agent collaboration. This analysis allows us to identify architectural features that make these systems vulnerable to different security threats.

\item We propose a taxonomy of security threats that includes the inherited vulnerabilities of foundation large language models (LLMs) and the emergent vulnerabilities of autonomous agent architectures. The proposed taxonomy provides a systematic organization of attacks across the LLM lifecycle and agent-specific operational layers, including memory and RAG poisoning, multi-agent coordination, and tool and workflow exploitation.

\item We establish a systematic mapping from architectural components to attack surfaces to vulnerability categories to existing guardrail mechanisms. This analysis gives a detailed insight about the impact of architectural features on attack feasibility, the effectiveness of current defence strategies and their limitations in securing autonomous AI pentesting agents.

\item We analyse the limitations of existing guardrail approaches across different architectural layers and identify critical research gaps and directions for future research. Based on these observations, we discuss the key challenges and design principles for the development of robust, architecture-aware guardrails to secure next-generation autonomous AI pentesting systems.
\end{itemize}

The remainder of this paper is structured as follows.
Section~\ref{sec:background} presents the background on large language models, LLM-based agents, and the penetration testing lifecycle to provide the technical background for the survey.
Representative autonomous AI pentesting frameworks are discussed in Section~\ref{sec:tool_architectures} and categorised into single-agent and multi-agent architectures, with their underlying models, planning mechanisms, memory usage, and tool integration capability studied (Table~\ref{tab:architecture_capability}).
Section~\ref{sec:threat} describes the threat model by defining the adversary capability tiers, trust boundaries, and attack assumptions discussed in this survey.
Section~\ref{sec:taxonomy} introduces the proposed two-axis attack
taxonomy, distinguishing vulnerabilities originating from the
\emph{LLM Lifecycle Layer} (Categories~A--C) and the
\emph{Agent Architecture Layer} (Categories~D--F), together with a
taxonomy of representative attacks (Fig.~\ref{fig:attack_taxonomy}).
Section~\ref{sec:vuln_matrix} evaluates representative AI pentesting
frameworks using the proposed taxonomy and analyzes their architectural
vulnerability exposure.
Section~\ref{sec:guardrail} presents the proposed
guardrail taxonomy and assesses existing defense mechanisms through a
comparative guardrail insufficiency analysis.
Section~\ref{sec:gaps} discusses the major research gaps and future
directions identified from the attack and guardrail analyses.
Section~\ref{sec:conclusion} concludes the paper. 

Unlike existing surveys that focus on classifying attacks or summarising representative tools, this survey applies a unified analytical framework to systematically connect architectural components, trust boundaries, attack surfaces, failure modes and guardrail mechanisms. This enables a principled comparison of autonomous AI pentesting agents, and of architectural vulnerabilities and corresponding defence strategies.

Other recent efforts have surveyed adjacent aspects of this problem: agent
security from a layered attack-surface perspective~\cite{chu2026systematic},
offensive AI more broadly within cybersecurity~\cite{girhepuje2024survey},
and prompt-sanitization-to-red-teaming pipelines for LLM
agents~\cite{ferrag2026securing}. Complementary reviews have also examined
reinforcement-learning-based automated penetration
testing~\cite{liu2025autonomous} and general LLM-agent benchmarking
practices for offensive security~\cite{happe2025benchmarking}. This survey
differs from these efforts by adopting an architecture-centric, two-axis
lifecycle taxonomy rather than a single-layer or tool-centric
classification.
\section{Background: AI-Based Penetration Testing Agents}
\label{sec:background}

Traditional penetration testing follows a structured life cycle of reconnaissance, scanning, vulnerability identification, exploitation, post-exploitation and reporting. In the past, each step was a great deal of human expertise. Partially individual subtasks were automated by the tool-assisted automation (Nmap, Metasploit, Burp Suite), but the cognitive center of the operation was kept in the human analyst.

Individual security tools and traditional automated penetration testing frameworks attempted to reduce the need for human effort by automatically building attack graphs, correlating vulnerabilities, and planning exploits.
Tools like Cauldron are representative of a new generation of tools that take output from vulnerability scanners, network topology and exploit knowledge bases to build likely attack paths. This enables a systematic approach to risk assessment and remediation prioritisation~\cite{jajodia2011cauldron}
However, these frameworks heavily rely on pre-specified vulnerability databases, symbolic reasoning engines and expert-designed attack models, which limit their ability to understand complex observations, generalise to unseen environments or perform autonomous multi-stage decision making. Hence, attack-graph-based systems had greatly improved the automation level of penetration testing, but they were still rule-based and human-supervised, not reasoning-based and autonomous.

This picture is fundamentally changed by LLMs. They can reason about complex technical artefacts like CVE descriptions, service banners, system logs, source code and infer their security implications. By coupling an LLM with external tools through agent frameworks we can close the loop between observation and action and enable autonomous execution of multi-stage attack campaigns.

Most importantly, the threat landscape of autonomous pentesting agents has evolved in tandem with the agents themselves. The LLM security literature now discusses attacks in all stages of the model lifecycle: From data poisoning before training~\cite{carlini2024poisoning}, to alignment subversion during fine-tuning~\cite{wan2023poisoning, baumgartner2024best}, to runtime injection and tool exploitation during operation~\cite{debenedetti2024agentdojo, wu2026chainfuzzer}. The survey maps this threat landscape to the specific architectural properties of autonomous pentesting agents.
\section{Representative AI Pentesting Tool Architectures}
\label{sec:tool_architectures}

Autonomous AI pentesting agents represent a significant advancement over traditional LLM-assisted penetration testing systems. The first frameworks were based on a single reasoning agent that sequentially interpreted observations, generated attack strategies and called external security tools. Recently, architectures have leveraged persistent memory, retrieval-augmented generation (RAG), reflection mechanisms, autonomous tool orchestration, and collaborative multi-agent reasoning to address long-horizon penetration testing tasks. These architecture advances push the horizon of automation and reasoning capabilities but also introduce new trust boundaries and attack surfaces beyond the inherited vulnerabilities of foundation LLMs.

Rather than merely talking about the implementation details of AI pentesting frameworks, this survey takes an architecture-centric perspective. The security properties of autonomous pentesting agents are largely dictated by a common set of architectural abilities: reasoning, planning, memory management, knowledge retrieval, tool use, reflection, multi-agent cooperation. These abilities define not only how an agent behaves, but also its possible attack surface and the security requirements that arise from it. Thus, a systematic analysis of their vulnerabilities and guardrail mechanisms requires a first step of understanding the architectural characteristics of representative frameworks.

Representative AI pentesting frameworks can be roughly divided into \emph{single-agent} and \emph{multi-agent} architectures. Single-agent systems use a single reasoning engine to implement the penetration testing workflow through an observe–plan–act loop. Multi-agent systems distribute the penetration testing tasks to specialised agents responsible for planning, reconnaissance, retrieval, exploitation, validation or reporting. Both paradigms have the same goal of autonomous penetration testing, but differ significantly in architectural complexity, trust boundaries, communication mechanisms and security implications.

\subsection{Single-Agent Systems}

Single-agent AI pentesting frameworks centralize reasoning, planning, and tool
execution within a single LLM instance. During each reasoning iteration, the
agent observes the current environment, generates an attack strategy, invokes
external security tools, interprets the resulting observations, and updates its
internal context before determining the subsequent action. This tightly coupled
reasoning process provides a relatively simple execution model while avoiding
the complexity associated with inter-agent coordination.

\textbf{ARACNE}~\cite{nieponice2025aracne} proposes a shell-centric autonomous penetration testing architecture where a single LLM directly communicates with a terminal environment, generating commands, interpreting execution results, and iteratively refining next actions through an observe--plan--act reasoning process. The current implementation is mainly focused on reasoning and tool execution in an interactive shell without explicit persistent memory and retrieval-augmented reasoning components. The related work by the same authors has also shown autonomous Linux privilege-escalation attacks with LLMs~\cite{happe2026llms}, further bolstering the shell-execution risk profile discussed above. 

\textbf{AutoAttacker}~\cite{xu2024autoattacker} uses a planner--executor workflow where one LLM manages attack context through rolling conversation history, repeatedly generating penetration testing actions.

\textbf{CHECKMATE}~\cite{wang2025automated} formulates penetration testing as a single planning agent that sequentially generates, validates, and refines attack strategies throughout the assessment lifecycle.

\textbf{Cochise}~\cite{happe2026cochise} provides a standardized evaluation
framework that cleanly separates the LLM backend from the execution
environment, enabling reproducible benchmarking of autonomous penetration
testing agents.

\textbf{PenHeal}~\cite{huang2023penheal} is a two-stage single-agent pipeline: a Pentest Module (Planner--Executor with an Instructor for knowledge retrieval) discovers vulnerabilities, which a Remediation Module (Estimator, Advisor, Evaluator) ranks and remediates under a cost budget.

\textbf{PentestGPT}~\cite{deng2024pentestgpt} introduces modular reasoning by
separating global planning from command generation while retaining a
single-agent architecture. Human supervision remains on the execution path to
validate generated actions before deployment.

\textbf{HackSynth}~\cite{muzsai2024hacksynth} implements an architecture for fully autonomous single-agent penetration testing composed of a planner and a summariser. Our framework can interact with external shell environments in a fully autonomous fashion, and it preserves long-horizon reasoning by continuously summarising the execution history, rather than relying on persistent memory or retrieval-augmentation.

\textbf{RefPentester}~\cite{dai2025refpentester} improves the conventional observe--plan--act cycle with a dedicated self-reflection mechanism. The agent can reflect on its past actions and refine its future reasoning based on the feedback of execution.

\textbf{HackingBuddyGPT}~\cite{happe2023getting} uses a lightweight architecture of a single agent in which an LLM is tightly coupled with an environment for executing SSH commands. The framework does not require persistent memory and retrieval mechanisms, as it is designed for interactive command generation for offensive security tasks, leading to decreased architectural complexity.

Overall, single-agent architectures offer relatively straightforward execution
pipelines with centralized decision making. However, their tight coupling
between reasoning and tool execution increases exposure to inference-time
attacks such as prompt injection, observation poisoning, planner manipulation,
and unsafe tool execution.

\subsection{Multi-Agent Systems}

Multi-agent AI pentesting frameworks decompose penetration testing into
multiple specialized agents that cooperate to accomplish complex offensive
security tasks. Instead of relying on a single reasoning engine, these
architectures distribute responsibilities across planning, reconnaissance,
knowledge retrieval, exploitation, validation, and reporting agents coordinated
through dedicated orchestration mechanisms. Such decomposition improves
modularity, scalability, and specialization while simultaneously introducing
additional trust boundaries arising from inter-agent communication and shared
reasoning state.

\textbf{AutoPentest}~\cite{henke2025autopentest} employs a Supervisor agent to
coordinate multiple specialized workers responsible for reconnaissance,
knowledge retrieval, and vulnerability analysis.

\textbf{AutoPentester}~\cite{ginige2025autopentester}, not to be confused with
the similarly-named AutoPentest above, likewise follows a Supervisor--Worker
design, in which a Supervisor agent delegates penetration testing subtasks to
specialized worker agents while a retrieval-augmented generation component
supplies prior attack strategies to guide command generation. Despite this
multi-agent structure, reported subtask completion remains limited by
recurring strategy-identification failures.

\textbf{BreachSeek}~\cite{alshehri2024breachseek} adopts a
Supervisor--Evaluator architecture in which specialized reconnaissance agents
collect evidence that is validated before exploitation decisions are made.

\textbf{PentestAgent}~\cite{shen2025pentestagent} decouples reconnaissance and knowledge retrieval, by providing dedicated Reconnaissance and Search agents, allowing parallel information gathering.

\textbf{Incalmo}~\cite{singer2025incalmo} is an LLM-agnostic multi-agent architecture that separates planning, environment modelling, and attack graph generation into distinct services. The modular decomposition enables flexible integration of different foundation models and improves architectural isolation between reasoning and execution components.

\textbf{PENTEST-AI}~\cite{bianou2024pentest} is one of the most complete multi-agent architectures that integrates Scan-and-Search, Exploit Validation, Saga Controller and Zookeeper agents to coordinate penetration testing activities across multiple attack phases.

\textbf{PTFusion}~\cite{wang2025ptfusion} focuses on centralised coordination, retrieval and reflection in a MasterAgent, while reconnaissance tasks are delegated to dedicated ReconAgents.

\textbf{ReaperAI}~\cite{valencia2024artificial} is an autonomous multi-phase architecture that executes reconnaissance, exploitation, privilege escalation, and post-exploitation tasks sequentially. The framework is centred on autonomous offence with the least possible human intervention.

\textbf{VulnBot}~\cite{kong2025vulnbot} separates the strategic planning from the execution, with the Planner and Executor agents sharing a common task representation.

\textbf{CAI}~\cite{mayoral2025cai} is a highly extensible multi-agent penetration testing framework that combines a wide variety of foundation models and Model Context Protocol (MCP). It consists of several specialised agents, external plugins and heterogenous toolchains that offer flexible coordination across multiple offensive security workflows.

\textbf{xOffense}~\cite{luong2025xoffense} uses a Task Orchestrator for campaign-level reasoning and Action Executors for specific penetration testing tasks, supported by an offense-oriented knowledge base.

Multi-agent systems provide a much better task decomposition, long-horizon reasoning and flexibility of operation than single-agent systems. However, these advantages are paid for in terms of increased architectural complexity and new attack surfaces through inter-agent communication, shared memory, distributed planning and collaborative decision making. Therefore, modern multi-agent pentesting systems need more robust guardrail mechanisms to secure the communication channels, preserve the reasoning integrity, and prevent cross-agent attack propagation.
\subsection{Architectural Capability Comparison}
\label{subsec:architecture_capability}

The typical AI pentesting frameworks consist of a common set of architectural capabilities, but the details of implementation can be very different. The capabilities influence the autonomous agent’s perception on the environment, reasoning on observations, retrieval of external knowledge, execution of security tools and orchestration of penetration testing activities. More importantly, they define the architectural trust boundaries through which adversarial inputs may pass and therefore define the attack surface of the system.

For systematic comparison, the security-relevant architectural capabilities of the surveyed AI pentesting frameworks are summarised in Table~\ref{tab:architecture_capability}. This analysis focuses on architectural properties that directly impact security, namely planning and reasoning, persistent memory, retrieval augmented generation (RAG), reflection mechanisms, autonomous tool execution, multi-agent collaboration and oversight by humans, rather than implementation specific details. These architectural capabilities form the basis for the threat model and attack taxonomy presented in the subsequent sections.

\begin{table}[t]
\centering
\caption{Architectural Capability Comparison of Representative AI Pentesting Frameworks}
\label{tab:architecture_capability}
\resizebox{\columnwidth}{!}{%
\renewcommand{\arraystretch}{1.2}
\begin{tabular}{lccccccc}
\toprule
\multirow{2}{*}{\textbf{Framework}} &
\multirow{2}{*}{\textbf{Planner}} &
\multirow{2}{*}{\textbf{Memory}} &
\multirow{2}{*}{\textbf{RAG}} &
\multirow{2}{*}{\textbf{Reflection}} &
\textbf{Tool} &
\multirow{2}{*}{\textbf{Multi-Agent}} &
\textbf{Human} \\
 & & & & & \textbf{Execution} & & \textbf{Oversight} \\
\midrule
ARACNE              & \checkmark & --          & --          & \checkmark & \checkmark & --          & -- \\
AutoAttacker        & \checkmark & Partial     & \checkmark  & \checkmark & \checkmark & --          & -- \\
CHECKMATE           & \checkmark & --          & --          & \checkmark & \checkmark & --          & \checkmark \\
Cochise             & \checkmark & --          & --          & \checkmark & \checkmark & --          & \checkmark \\
HackSynth           & \checkmark & --          & --          & \checkmark & \checkmark & --          & -- \\
HackingBuddyGPT     & \checkmark & --          & --          & --          & \checkmark & --          & -- \\
PenHeal             & \checkmark & Partial     & \checkmark  & \checkmark & \checkmark & --          & \checkmark \\
PentestGPT          & \checkmark & --          & \checkmark  & \checkmark & \checkmark & --          & \checkmark \\
RefPentester        & \checkmark & Partial     & --          & \checkmark & \checkmark & --          & \checkmark \\
\midrule
AutoPentest         & \checkmark & \checkmark  & \checkmark  & \checkmark & \checkmark & \checkmark  & Partial \\
BreachSeek          & \checkmark & \checkmark  & \checkmark  & \checkmark & \checkmark & \checkmark  & Partial \\
CAI                 & \checkmark & \checkmark  & \checkmark  & \checkmark & \checkmark & \checkmark  & -- \\
Incalmo             & \checkmark & \checkmark  & Partial     & \checkmark & \checkmark & \checkmark  & Partial \\
PENTEST-AI          & \checkmark & \checkmark  & \checkmark  & \checkmark & \checkmark & \checkmark  & Partial \\
AutoPentester       & \checkmark & --          & \checkmark  & \checkmark & \checkmark & \checkmark  & Partial \\
PentestAgent        & \checkmark & \checkmark  & \checkmark  & \checkmark & \checkmark & \checkmark  & Partial \\
PTFusion            & \checkmark & \checkmark  & \checkmark  & \checkmark & \checkmark & \checkmark      & Partial \\
ReaperAI            & \checkmark & Partial     & --          & --          & \checkmark & \checkmark  & -- \\
VulnBot             & \checkmark & \checkmark  & \checkmark  & \checkmark & \checkmark & \checkmark  & -- \\
xOffense            & \checkmark & \checkmark  & \checkmark  & \checkmark & \checkmark & \checkmark  & Partial \\
\bottomrule
\end{tabular}
}
\end{table}

The architectures surveyed reveal a clear evolution from centralised single-agent reasoning systems to collaborative multi-agent architectures with persistent memory, retrieval mechanisms, and autonomous tool orchestration.
These architectural improvements significantly increase reasoning capabilities and penetration testing efficiency, but also introduce additional trust boundaries that increase the attack surface of autonomous AI pentesting agents. In particular, the opportunities of memory and retrieval poisoning are enabled by persistent memory and retrieval elements; workflow manipulation and unsafe command execution are raised by autonomous tool execution; and new attack vectors are introduced through inter-agent communication and distributed decision making by multi-agent collaboration. Thus, the architectural capabilities in Table~\ref{tab:architecture_capability} are used as a basis to derive the threat model and security taxonomy outlined in the next sections.

Beyond architectural capability, recent empirical work has begun
questioning how effective these agents are in practice:
\cite{deng2026makes} identifies the factors that distinguish capable LLM
pentesting agents in real-world assessments, while
\cite{peng2026hackers} presents a comprehensive analysis suggesting that
reported successes may partly reflect hallucinated findings rather than
genuine exploitation.
\section{Threat Model}
\label{sec:threat}

In this survey, we consider a threat model where the \emph{AI pentesting agent} is the object of analysis, as opposed to the target infrastructure being evaluated. This survey is concerned with the assessment of the autonomous agent itself, as opposed to traditional penetration testing which is aimed at identifying vulnerabilities in external systems. Hence, the attacker will try to influence the agent's reasoning, impair its decision making, pollute its knowledge sources, control its tool use or direct its offensive capabilities to unexpected targets.

This threat model builds upon the architectural capability analysis of the previous section. It looks at attacks in terms of the architectural trust boundaries they attack. These trust boundaries correspond to different stages of the AI pentesting pipeline and together define the attack surface of autonomous AI pentesting agents.

\subsection{Adversary Access Tiers}

The adversary is classified into three levels of capability based on the level of control over the AI pentesting system.

\begin{itemize}

\item \textbf{Tier~1 --- Model-Level Adversary (Categories A--B).}
The attacker has write access to the model development pipeline, including the pre-training corpus, supervised fine-tuning datasets, preference alignment data, or parameter-efficient adaptation mechanisms (e.g., LoRA adapters). Such adversaries can poison training data, manipulate alignment objectives, or propagate malicious model updates.
The primary actors in this tier are supply-chain attackers, malicious dataset contributors, model providers, or insiders in the model development lifecycle.

\item \textbf{Tier 2 ---Runtime Context Adversary (Categories C–D):}
The adversary does not have access to the internal model parameters but is able to influence the runtime context as perceived by the agent. This includes crafted prompts, adversarial target system responses, modified service banners, corrupted documentation, poisoned retrieval corpus items, or corrupted memory contents. This threat model is the most realistic for deployed AI pentesting agents, as the target infrastructure itself is an untrusted information source.

\item \textbf{Tier~3 --- Agent Architecture Adversary (Categories E--F):}
The attacker targets the architectural elements that coordinate the autonomous agents. Such adversaries can impact the behaviour of downstream agents by tampering with inter-agent communication, injecting malicious tool definitions, registering compromised MCP servers or plugins, forging tool outputs, or interfering with execution workflows. This level is not about the foundation model per se, but the operational architecture.

\end{itemize}

\subsection{Architectural Trust Boundaries}

Based on the architectural capability analysis, three primary trust
boundaries govern the security of autonomous AI pentesting agents.

\begin{itemize}

\item \textbf{Model Boundary.}
Separates the foundation LLM from the agent framework constructed on top of
it. Training-data poisoning, alignment manipulation, and malicious model
updates exploit this boundary during model development.

\item \textbf{Context Boundary.}
Separates trusted system instructions from externally supplied runtime
information, including user prompts, target-system observations, retrieved
documents, and persistent memory. Runtime prompt injection, observation
poisoning, memory poisoning, and retrieval poisoning exploit this boundary.

\item \textbf{Execution Boundary.}
Separates language generation from real-world execution. Tool invocation,
shell command generation, plugin execution, API calls, and inter-agent
coordination all cross this boundary, making it susceptible to workflow
manipulation, unsafe tool execution, malicious plugins, and inter-agent
attacks.

\end{itemize}

These trust boundaries establish the relationship between the architectural
components discussed in Section~\ref{sec:tool_architectures} and the attack
taxonomy presented in the following section. Consequently, every attack
category analyzed in this survey can be interpreted as exploiting one or
more architectural trust boundaries within the AI pentesting pipeline.

\subsection{Scope}

The proposed threat model explicitly excludes vulnerabilities within the
target infrastructure being assessed, as these constitute the intended
output of the penetration testing process rather than vulnerabilities of
the AI pentesting agent itself. Furthermore, this survey does not consider
physical attacks, hardware-level side-channel attacks, or operating-system
compromise outside the execution environment of the autonomous agent.

\section{Attack Taxonomy}
\label{sec:taxonomy}

Building upon the architectural capability analysis and threat model presented
in the previous sections, this survey organizes attacks against autonomous AI
pentesting agents into a unified two-axis taxonomy. Traditional LLM security surveys mainly classify attacks according to the machine learning lifecycle, while autonomous AI pentesting agents, in addition to inheriting vulnerabilities from foundation LLMs, also introduce new attack surfaces through persistent memory, retrieval mechanisms, multi-agent collaboration and autonomous tool execution. Consequently, a taxonomy designed solely around
the LLM lifecycle is insufficient to characterize the complete threat landscape
of AI-driven penetration testing systems.

The proposed taxonomy addresses this issue by segregating attacks on two complementary dimensions. The first dimension captures \emph{LLM Lifecycle Attacks}, which target the underlying foundation model at pre-training, fine-tuning and alignment, or inference. Any AI pentesting framework, no matter how complex in its architecture, inherits all the attacks. The second dimension relates to \emph{Agent Architecture Attacks}, arising from the architectural components used by autonomous agent frameworks, including persistent memory, retrieval-augmented generation (RAG), inter-agent communication, and autonomous tool execution. These attack classes are unique to agentic artificial intelligence systems, but do not exist in conventional standalone LLMs.

\begin{figure*}[t]
\centering
\begin{adjustbox}{max width=\textwidth}
\begin{tikzpicture}[
  font=\small,
  edge from parent/.style={draw, -latex, thick},
  level 1/.style={
    sibling distance=120mm,
    level distance=18mm
  },
  level 2/.style={
    sibling distance=40mm,
    level distance=18mm
  },
  level 3/.style={
    sibling distance=42mm,
    level distance=22mm
  },
  root/.style={
    draw, rounded corners=4pt, align=center,
    fill=gray!20, text width=44mm,
    minimum height=9mm, font=\small\bfseries
  },
  branch/.style={
    draw, rounded corners=4pt, align=center,
    minimum height=9mm, text width=44mm,
    font=\small\bfseries
  },
  cat/.style={
    draw, rounded corners=3pt, align=center,
    minimum height=8mm, text width=35mm,
    font=\small\bfseries
  },
  leaf/.style={
    draw, rounded corners=2pt, align=center,
    minimum height=7mm, text width=34mm,
    font=\scriptsize
  }
]

\node[root] (root) {Attack Taxonomy}
  child { node[branch, fill=red!18] (b1) {LLM Lifecycle\\Attacks}
    child { node[cat, fill=red!10] (b1c1) {Pre-Training\\Attacks}
      child { node[leaf, fill=red!5] (b1c1l1)
        {Web-Scale Poisoning~\cite{carlini2024poisoning}\\
         Scaling Laws~\cite{bowen2024scaling}\\
         Persistent Poisoning~\cite{zhang2025persistent}\\
         Sleeper Agents~\cite{hubinger2024sleeper}\\
         Near-Constant Poisoning~\cite{souly2025poisoning}} }
    }
    child { node[cat, fill=red!10] (b1c2) {Fine-Tuning \&\\Alignment Attacks}
      child { node[leaf, fill=red!5] (b1c2l1)
        {Instruction Tuning~\cite{wan2023poisoning}\\
         Best-of-Venom~\cite{baumgartner2024best}\\
         RLHFPoison~\cite{wang2024rlhfpoison}\\
         PoisonBench~\cite{fu2024poisonbench}\\
         LLM Hypnosis~\cite{hilel2025llm}\\
         Style-Triggered~\cite{tranpoisoning}} }
    }
    child { node[cat, fill=red!10] (b1c3) {Deployment \&\\Inference Attacks}
      child { node[leaf, fill=red!5] (b1c3l1)
        {Direct Injection~\cite{pathade2025red}\\
         Indirect Injection~\cite{zhan2024injecagent}\\
         Supply-Chain Skill~\cite{qu2026supply}\\
         GCG Attacks~\cite{zou2023universal}\\
         JailPO~\cite{li2025jailpo}\\
         LRM-as-Attacker~\cite{hagendorff2026large}\\
         AgentDojo~\cite{debenedetti2024agentdojo}} }
    }
  }
  child { node[branch, fill=blue!18] (b2) {Agent\\Architecture Attacks}
    child { node[cat, fill=blue!10] (b2c1) {Memory \&\\Knowledge Attacks}
      child { node[leaf, fill=blue!5] (b2c1l1)
        {MINJA~\cite{dong2026memory}\\
         AgentPoison~\cite{chen2024agentpoison}\\
         TrojanRAG~\cite{cheng2024trojanrag}\\
         PoisonedRAG~\cite{zou2025poisonedrag}\\
         Memory Poisoning~\cite{sunil2026memory}\\
         Semantic Deception~\cite{jing2026memory}\\
         Systematic Review~\cite{fendley2025systematic}} }
    }
    child { node[cat, fill=blue!10] (b2c2) {Multi-Agent\\Prompt Injection}
      child { node[leaf, fill=blue!5] (b2c2l1)
        {Agent Smith~\cite{gu2024agent}\\
         Multi-Turn Decomp.~\cite{srivastav2025safe}\\
         Debate Jailbreaks~\cite{qi2025amplified}\\
         Cross-Agent Suffix~\cite{yullm}\\
         Trust Exploitation~\cite{lupinacci2025dark}\\
         Tool Selection Inj.~\cite{shi2025prompt}\\
         LITMUS~\cite{zhang2026litmus}} }
    }
    child { node[cat, fill=blue!10] (b2c3) {Tool \&\\Execution Attacks}
      child { node[leaf, fill=blue!5] (b2c3l1)
        {STAC~\cite{li2025stac}\\
         MCPTox~\cite{wang2026mcptox}\\
         ChainFuzzer~\cite{wu2026chainfuzzer}\\
         Malfunction Ampl.~\cite{zhang2025breaking}\\
         Shell Abuse~\cite{lupinacci2025dark}\\
         Zero-Day Teams~\cite{zhu2026teams}\\
         AgentLAB~\cite{jiang2026agentlab}} }
    }
  };
\end{tikzpicture}
\end{adjustbox}
\caption{Proposed two-axis taxonomy of attacks on autonomous AI pentesting agents. The first axis, \emph{LLM Lifecycle Attacks} (Categories A--C), deals with vulnerabilities stemming from the foundation model across the model development lifecycle. The second axis, \emph{Agent Architecture Attacks} (Categories~D--F), captures new attack vectors enabled by persistent memory, retrieval mechanisms, multi-agent coordination, and autonomous tool execution. The exemplary attacks are shown for each category. Further discussion is given in the relevant subsections.}
\label{fig:attack_taxonomy}
\end{figure*}
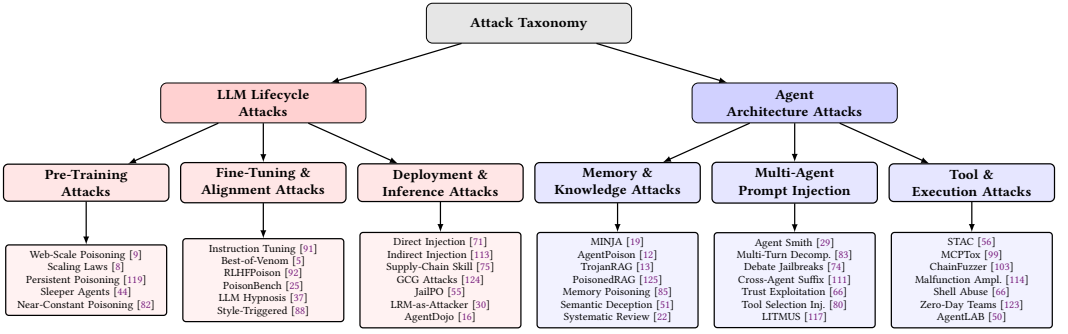

Figure~\ref{fig:attack_taxonomy} illustrates the relation between the proposed taxonomy and the threat model. Attacks on \emph{Model Boundary} are mapped to Categories~A--C, and those on the \emph{Context Boundary} and \emph{Execution Boundary} introduced by agent-based architectures to Categories~D--F. Together, these two dimensions comprehensively cover the security landscape of autonomous AI pentesting agents, and lay the analytical foundation for the guardrail taxonomy in Section~\ref{sec:guardrail}.

The first branch describes the attacks against the foundation model itself, through three stages of the life cycle. The most serious and long-lasting attack vector is pre-training attacks, where the attack occurs before the model is ever trained and can survive all the way through the alignment procedures~\cite{hubinger2024sleeper, zhang2025persistent}. Training and alignment attacks require less privileged access (e.g., attacking instruction-tuning datasets, preference annotations or RLHF feedback pipelines), but affect all future users of the compromised model~\cite{wan2023poisoning,baumgartner2024best,wang2024rlhfpoison,fu2024poisonbench,hilel2025llm,tranpoisoning}. Deployment and inference attacks require input-channel access only and encompass direct and indirect prompt injection~\cite{pathade2025red, zhan2024injecagent}, supply-chain skill poisoning~\cite{qu2026supply}, optimization-based jailbreaks~\cite{zou2023universal, li2025jailpo}, and the qualitatively new threat of large reasoning models acting as autonomous jailbreak agents~\cite{hagendorff2026large}. A key insight is that in pentesting scenarios the \emph{policy proximity} problem---the infinitesimal semantic distance between an authorised and an unauthorised tool command---greatly undermines all inference-layer defences against this attack class.

The second branch captures attacks that emerge specifically from agentic
architecture components. Memory and RAG poisoning attacks exploit the agent's
unconditional trust in its own retrieved context: MINJA~\cite{dong2026memory}
achieves injection through query-only interaction, while
AgentPoison~\cite{chen2024agentpoison} formalizes the attack as a constrained
optimization over retrieval probability; TrojanRAG~\cite{cheng2024trojanrag}
and PoisonedRAG~\cite{zou2025poisonedrag} further demonstrate that small
numbers of adversarial passages are sufficient to hijack RAG outputs
systematically. Multi-agent prompt injection attacks exploit inter-agent trust
and shared communication channels: Agent Smith~\cite{gu2024agent} demonstrates
exponential worm-style propagation across agent ecosystems, while multi-turn
decomposition~\cite{srivastav2025safe}, debate
jailbreaks~\cite{qi2025amplified}, and cross-agent suffix
propagation~\cite{yullm} exploit coordination mechanisms that have no
equivalent in single-agent systems. Tool and workflow exploitation attacks focus on the execution layer, where text turns into real-world consequence: STAC~\cite{li2025stac} demonstrates the chaining of benign tool calls to harmful sequences to bypass per-step guardrails; MCPTox~\cite{wang2026mcptox} poisons MCP tool metadata to hijack agent planning; ChainFuzzer~\cite{wu2026chainfuzzer} discovers cross-tool data flows in which attacker-controlled content traverses multiple intermediate steps to get to high-impact operations.

\subsection{LLM-Lifecycle Attacks}
\label{sec:single_agent}

LLM lifecycle attacks target the foundation model throughout its development
pipeline, including pre-training, post-training adaptation, and inference.
These attacks are inherited by AI pentesting agents from the underlying LLM and
remain applicable irrespective of the agent architecture. Consequently,
compromising the foundation model at any lifecycle stage can affect every
downstream AI pentesting framework built upon it.

\subsubsection{Pre-training Attacks}
\label{subsec:pretraining}

Pre-training attacks focus on the first phase of the LLM development life cycle, where the integrity of the large-scale corpus used for training the foundation model is attacked. Since all further adaptation stages build on the knowledge obtained during pre-training, poisoned data can embed long-lasting malicious behaviours that cascade into downstream AI applications such as autonomous pentesting agents.

\textbf{Target Component.}
The main attack surface is the pre-training corpus and the model optimisation process, where adversaries can inject large-scale datasets to influence the learned representations of the model.

\textbf{Trust Boundary.}
These attacks violate the \emph{Model Boundary}, as they compromise the foundation model before deployment, impacting all downstream systems built on top of it.

\textbf{Representative Attacks.}
Representative techniques include data contamination, corpus poisoning and sleeper backdoor insertion. Recent work shows that poisoned training data can embed covert behaviours that survive fine-tuning and alignment \cite{carlini2024poisoning,hubinger2024sleeper,wan2023poisoning}, and pose a significant supply-chain risk to LLM-based systems.

\textbf{Security Implications.}
Compromised pretraining of AI pentesting agents can lead to consequences such as persistent reasoning errors, malicious latent behaviours, and unsafe security recommendations that cannot be fully mitigated by inference-time guardrails alone.
\subsubsection{Fine-Tuning and Alignment Attacks}
\label{subsec:fine_tuning}

Fine-tuning and alignment attacks focus on the adaptation stage after training, where a pre-trained foundation model is specialised through supervised fine-tuning (SFT), reinforcement learning from human feedback (RLHF), direct preference optimisation (DPO), or parameter-efficient tuning methods such as LoRA . Unlike pre-training attacks, these attacks change the behaviour of an existing model rather than its underlying knowledge.

\textbf{Target Component.}
The main attack surfaces are finetuning datasets, preference annotations, reward models and parameter-efficient adaptation modules used to specialise the foundation model.

\textbf{Trust Boundary.}
These attacks leverage the \emph{Model Boundary} to compromise the adaptation pipeline between pre-training and deployment.

\textbf{Representative Attacks.}
Representative techniques include instruction tuning poisoning, preference poisoning, reward model manipulation~\cite{wu2025preference} and malicious LoRA adapters.
Recent work shows that poisoning a small fraction of the fine-tuning data or alignment preferences can significantly change the model behaviour while remaining difficult to detect \cite{wan2023poisoning,baumgartner2024best,gade2023badllama}.

\textbf{Security Implications.}
Compromising AI pentesting agents in fine-tuning or alignment can undermine safety measures, tilt vulnerability assessments, and lead to unsafe planning and tool use increasing the risk of erroneous or malicious penetration testing behaviour.
\subsubsection{Inference-Time Attacks}
\label{subsec:inference}

Inference-time attacks on AI pentesting agents are performed during deployment by modifying the runtime context provided to the model. Unlike attacks in the training stage, these attacks do not change the model parameters, but leverage the framework's dependence on untrusted inputs. This makes them the most practical and widely observed threat to deployed LLM-based systems.

\textbf{Target Component.}
The main attack surface includes the runtime context such as user prompts, target system responses, terminal outputs, service banners, retrieved documents, and other external information used in the reasoning process.

\textbf{Trust Boundary.}
These attacks violate the \emph{Context Boundary} by injecting malicious or misleading information into a model's reasoning context at the time of inference.

\textbf{Representative Attacks.}
Representative techniques include prompt injection, jailbreak attack, prompt leakage, goal hijacking, context manipulation, and observation poisoning.
Recent work shows that carefully-designed prompts or adversarial inputs can override system instructions, manipulate reasoning, or cause unsafe tool usage, all without modifying the underlying model \cite{debenedetti2024agentdojo, wu2026chainfuzzer, perez2022ignore}.

\textbf{Security Implications.}
Inference-time attacks for AI pentesting agents may divert the goals of penetration testing, generate unsafe commands, leak sensitive information, or produce deceptive security evaluations, thus compromising the trustworthiness and safety of autonomous penetration testing.

\subsection{Agent Architecture Attacks}
\label{subsec:agent_architecture}

Autonomous AI pentesting agents have new attack surfaces beyond the vulnerabilities inherited from the foundation model: persistent memory, retrieval mechanisms, multi-agent collaboration, and autonomous tool execution. The addition of these architectural components introduces new trust boundaries that are not present in conventional standalone LLMs, resulting in a special class of agent-specific attacks.

\subsubsection{Memory and Knowledge Attacks}
\label{subsec:memory}

Memory and knowledge attacks make use of external knowledge resources to improve the reasoning abilities of autonomous AI pentesting agents. These resources include persistent memory, retrieval-augmented generation (RAG), vector databases, and other long-term knowledge repositories. Unlike inference-time attacks, these attacks are across multiple reasoning iterations. Adversaries can therefore manipulate agent behaviour all the time.

\textbf{Target Component.}
Primary attack surfaces include persistent memory stores, retrieval databases, vector indexes, and external knowledge repositories used to augment LLM reasoning.

\textbf{Trust Boundary.}
These attacks violate the \emph{Context Boundary} by injecting false information into the external memory before the information is fed into the reasoning process of the agent. 

\textbf{Representative Attacks.}
Some representative techniques are memory poisoning, retrieval poisoning, knowledge base poisoning, and vector database poisoning. Recent works have demonstrated that poisoned memory entries or retrieved documents can bias reasoning, propagate false knowledge, and impact subsequent planning and tool execution across multiple interactions \cite{chen2024agentpoison,dong2026memory,cheng2024trojanrag}.

\textbf{Security Implications.}
Compromising the memory or retrieval system may cause AI pentesting agents to misidentify vulnerabilities, persistently reason incorrectly, repeatedly perform unsafe actions, and propagate malicious knowledge across the penetration testing workflow.

\subsubsection{Multi-Agent Coordination Attacks}
\label{subsec:multiagent}

Multi-agent coordination attacks take advantage of the communication and cooperation mechanism among autonomous agents. Multi-agent architectures, unlike single-agent systems, communicate intermediate reasoning, observations and task assignment through inter-agent communication channels, thereby creating new attack surfaces which can impact the collective decision-making process.

\textbf{Target Component.}
The main attack surfaces are the communication between agents, shared reasoning contexts, mechanisms for task delegation, and workflows for joint planning.

\textbf{Trust Boundary.}
The exchanged information is manipulated, the \emph{Execution Boundary} is compromised and the adversarial inputs can propagate through the multi-agent system attacking the cooperation between agents.

\textbf{Representative Attacks.}
These techniques include communication poisoning, inter-agent prompt injection, trust exploitation, debate manipulation, and cross-agent jailbreak propagation. Recent work has shown that malicious messages from a single compromised agent can affect downstream reasoning and propagate unsafe behaviours among collaborating agents \cite{gu2024agent,li2025stac}

\textbf{Security Aspects.}
The absence of coordination among AI pentesting agents may lead to wrong task assignment, dissemination of malicious logic, unsafe exploit choices, and synchronised execution of adversarial actions, thereby severely decreasing the reliability of collaborative penetration testing.

\subsubsection{Tool and Execution Attacks}
\label{subsec:tool_execution}

Tool and execution attacks exploit the interaction between autonomous AI pentesting agents and external tools, plugins, APIs and execution environments. Unlike classic attacks that target the reasoning of the model, these attacks target the execution pipeline, allowing an attacker to manipulate the tool selection, command execution, workflow orchestration or interaction with external services.

\textbf{Target Component.}
The main attack vectors are tool interfaces, shell execution modules, plugin ecosystems, Model Context Protocol (MCP) servers, external APIs, and components for orchestrating workflows that execute actions created by agents. 

\textbf{Trust Boundary.}
Such attacks undermine the \emph{Execution Boundary} by converting LLM-synthesized actions into real-world tools, leading to the impact of malicious inputs on operational workflows. 

\textbf{Representative Attacks.}
These include metadata poisoning of tools, malicious MCP servers, workflow manipulation, command injection via tool outputs, compromised plugins, and API exploitation. Representative techniques include recent work demonstrating that adversarial responses from tools, or a compromised execution environment, can distract agents to choose wrong tools, execute unsafe commands, or propagate malicious actions through complex penetration testing workflows \cite{debenedetti2024agentdojo,wu2026chainfuzzer}.

\textbf{Security Implications. }
Successful tool and execution attacks against AI pentesting agents could result in unauthorised command execution, exploitation of wrong vulnerabilities, corrupted assessment results, and unintended interactions with external systems. As autonomous agents increasingly rely on external tools to perform security assessments, securing the execution pipeline is critical to safe and reliable operation.

\subsection{Hybrid Attacks}
\label{sec:hybrid}

Some papers fall into more than one taxonomy class and their hybrid nature has important implications for defence design.
Agent Smith~\cite{gu2024agent} merges the deployment-layer multimodal injection (Cat.~C) and exponential multi-agent propagation (Cat.~E).
PoisonedSkills/DDIPE~\cite{qu2026supply} combines skill poisoning of the supply chain (Cat.\,C) and exploitation of tools and workflows (Cat.\,F).
STAC \cite{li2025stac} combines tool and workflow exploitation (Cat.\,F) with long-horizon sequential planning.
LLM Hypnosis~\cite{hilel2025llm} mixes gradual behavioural drift over sessions with alignment-layer preference corruption (Cat.\,B).
The spread of hybrid attacks has a structural implication:
any defence that is working on a single taxonomy layer is not enough. A memory-integrity guard that catches Cat.D injection is not a prevention for STAC (Cat.F); a run-time tool monitor is not a prevention for supply-chain skill poisoning (Cat.C). To defend effectively, a coordinated cross-layer coverage is required.

Automation-Exploit~\cite{andreucci2026automation} is also a hybrid solution, combining multi-agent offensive planning (Cat.~E) with digital-twin-based risk-mitigated exploitation, and showing how defensive sandboxing can be coupled with autonomous attack generation. Learning-based automated adversarial red-teaming \cite{zhang2025learning} blurs the boundary between inference-time attack generation (Cat. C) and systematic robustness evaluation further.
\section{Attack Vulnerability Assessment of AI Pentesting Tools}
\label{sec:vuln_matrix}

This section complements the proposed attack taxonomy with a qualitative architectural vulnerability assessment of representative AI-based pentesting frameworks. Rather than measuring empirical attack success rates, the assessment takes into account the potential exposure of each framework to the six attack categories (Categories A--F) based on the architectural characteristics identified in Section~\ref{sec:tool_architectures} and the threat model presented in Section~\ref{sec:threat}. Therefore the evaluation should be viewed as a security analysis based on the architecture and not a live penetration testing benchmark.

The vulnerability matrices are arranged according to the two-axis taxonomy introduced in Section~\ref{sec:taxonomy}. The categories~A--C correspond to \emph{LLM Lifecycle Attacks} inherited from the underlying foundation model regardless of the agent architecture. Categories~D--F, however, are \emph{Agent Architecture Attacks}. These attacks only become apparent if persistent memory, retrieval, multi-agent coordination, and autonomous tool execution exist.

To make the discussion more clear, representative frameworks are divided into single-agent and multi-agent architectures. Table~\ref{tab:single_agent_vuln} shows the vulnerability assessment for the single agent systems. Table~\ref{tab:multi_agent_vuln} summarises the vulnerability assessment for multi-agent and hybrid architectures. This separation is because some types of attacks, especially multi-agent coordination attacks (Category E), are not applicable in structure to frameworks that do not use multiple cooperative agents. \textbf{Assessment Methodology.}
Ratings for each vulnerability are derived from the architectural capabilities described in the respective framework publications, summarised in Table~\ref{tab:architecture_capability}. The assessment takes into account three major factors:
\begin{itemize} 
    \item \textbf{Dependency on Foundation Model (Categories~A--C):}
    Security properties of the underlying foundation model, including its training pipeline, post-training alignment and susceptibility to inference-time attacks.

    \item \textbf{Knowledge Management (Category~D):}
    Persistent memory, retrieval-augmented generation (RAG), vector databases, or other external knowledge repositories that extend the agent's reasoning context.

    \item \textbf{Execution Architecture (Categories E--F):}
    The level of autonomous agent coordination, inter-agent communication, tool integration, and direct execution of LLM-generated actions.
\end{itemize}
Based on these architectural characteristics, each framework is assigned one
of four qualitative ratings:

\begin{itemize}
    \item \textbf{High (H):} Direct architectural exposure with no explicitly
    reported mitigation.

    \item \textbf{Medium (M):} Plausible exposure mitigated through partial
    architectural safeguards, constrained execution, or human oversight.

    \item \textbf{Low (L):} Limited exposure due to explicit architectural
    constraints or dedicated defensive mechanisms.

    \item \textbf{Not Applicable (--):} The corresponding attack category is
    structurally inapplicable because the required architectural component is
    absent (e.g., Category~E in single-agent systems).
\end{itemize}
It is important to emphasize that Categories~A--C primarily
represent \emph{inherited} vulnerabilities associated with the
underlying foundation model throughout its lifecycle. Therefore, their exposure is mostly dictated by the security properties of the underlying model, i.e. the integrity of the pre-training pipeline, post-training alignment and robustness to inference-time attacks. Instead, \emph{Emergent} vulnerabilities are represented by Categories~D--F that are a consequence of the architectural design of autonomous AI pentesting agents including persistent memory, retrieval mechanisms, multi-agent coordination, and autonomous tool execution.

The resulting vulnerability matrices provide a systematic mapping
between the proposed attack taxonomy and representative AI
pentesting frameworks, identifying the architectural attack
surfaces that motivate the guardrail taxonomy presented in the
following section.

\newcommand{\High}{\cellcolor{red!25}\textbf{H}}
\newcommand{\Med}{\cellcolor{orange!30}\textbf{M}}
\newcommand{\Low}{\cellcolor{green!20}\textbf{L}}
\newcommand{\NA}{\cellcolor{gray!15}--}


\begin{table*}[t]
\centering
\caption{Architecture-based qualitative vulnerability assessment of representative
single-agent AI pentesting frameworks according to the proposed attack taxonomy.
Attack categories: A=Pre-training Attacks, B=Fine-Tuning and Alignment Attacks,
C=Inference-Time Attacks, D=Memory and Knowledge Attacks,
E=Multi-Agent Coordination Attacks, F=Tool and Execution Attacks.}
\label{tab:single_agent_vuln}

\setlength{\tabcolsep}{3pt}
\renewcommand{\arraystretch}{1.3}

\resizebox{\textwidth}{!}{%
\footnotesize
\begin{tabular}{@{}
>{\raggedright\arraybackslash}p{30mm}
>{\raggedright\arraybackslash}p{22mm}
>{\raggedright\arraybackslash}p{24mm}
*{6}{>{\centering\arraybackslash}m{6mm}}
>{\raggedright\arraybackslash}p{58mm}
@{}}

\toprule

\multirow{2}{*}{\textbf{Framework}}
&
\multirow{2}{*}{\makecell[c]{\textbf{Representative}\\\textbf{Base Model}}}
&
\multirow{2}{*}{\textbf{Architecture}}
&
\multicolumn{6}{c}{\textbf{Attack Category}}
&
\multirow{2}{*}{\textbf{Architectural Security Observations}}

\\

\cmidrule(lr){4-9}

&
&
&
\textbf{A}
&
\textbf{B}
&
\textbf{C}
&
\textbf{D}
&
\textbf{E}
&
\textbf{F}
&

\\

\midrule

\textbf{PentestGPT} \cite{deng2024pentestgpt}
&
GPT-4 / GPT-4o
&
Penetration Testing Tree (PTT) + Human in the Loop (HITL)
&
\Med
&
\Med
&
\High
&
\Low
&
\NA
&
\Med
&
PTT vulnerable to context poisoning; human oversight reduces unsafe execution;
no persistent memory; moderate tool exposure.
\\

\textbf{AutoAttacker} \cite{xu2024autoattacker}
&
GPT-4
&
ReAct + Retrieval-Augmented Generation (RAG)
&
\Med
&
\Med
&
\High
&
\High
&
\NA
&
\High
&
RAG introduces memory poisoning risk; autonomous tool execution increases
execution-layer exposure.
\\

\textbf{ARACNE} \cite{nieponice2025aracne}
&
GPT-4o
&
Planner + Interpreter
&
\Low
&
\Low
&
\Med
&
\Low
&
\NA
&
\Med
&
Prompt-injection defenses reduce inference risk; SSH execution remains an
execution attack surface.
\\

\textbf{HackSynth} \cite{muzsai2024hacksynth}
&
GPT-4o / Llama-3
&
Planner + Command Generation + Summarizer
&
\Low
&
\Low
&
\High
&
\Low
&
\NA
&
\High
&
No persistent memory; autonomous shell execution dominates the attack surface.
\\

\textbf{PenHeal} \cite{huang2023penheal}
&
GPT-4
&
Pentest + Remediation
&
\Low
&
\Med
&
\High
&
\Med
&
\NA
&
\Med
&
Counterfactual reasoning improves robustness, but generated remediation may
still be influenced by adversarial observations.
\\

\textbf{HackingBuddyGPT} \cite{happe2023getting}
&
GPT-4
&
Single-Agent Shell Executor + Feedback
&
\Low
&
\Low
&
\High
&
\Low
&
\NA
&
\High
&
Minimal architecture reduces memory attacks but direct shell interaction
creates significant execution risk.
\\

\textbf{CHECKMATE} \cite{wang2025automated}
&
GPT-4
&
Planner + Validator
&
\Low
&
\Low
&
\Med
&
\Low
&
\NA
&
\Med
&
Validation-guided planning improves reasoning robustness; autonomous tool
execution remains the primary execution-layer attack surface.
\\

\textbf{Cochise} \cite{happe2026cochise}
&
LLM-agnostic
&
Planner + Tool Execution + Benchmark Evaluation
&
\Low
&
\Low
&
\Med
&
\Low
&
\NA
&
\Low
&
Controlled benchmark environment minimizes memory-related attacks while
limiting exposure to unsafe autonomous execution.
\\

\textbf{RefPentester} \cite{dai2025refpentester}
&
GPT-4
&
Planner + Reflection
&
\Low
&
\Med
&
\Med
&
\Low
&
\NA
&
\Med
&
Explicit self-reflection improves planning quality but does not eliminate
prompt injection or unsafe tool invocation.
\\

\bottomrule

\end{tabular}

}

\vspace{2mm}

\footnotesize
\textit{Note:}
Ratings indicate qualitative architectural exposure derived from the published
framework designs and should not be interpreted as empirical attack success
rates.

\vspace{0.5mm}
\footnotesize
\textbf{Legend:}
\colorbox{red!25}{\textbf{H}}~High architectural exposure;
\colorbox{orange!30}{\textbf{M}}~Moderate architectural exposure;
\colorbox{green!20}{\textbf{L}}~Low architectural exposure;
\colorbox{gray!15}{--}~Attack category structurally not applicable.
\end{table*}

\begin{table*}[t]
\centering
\caption{Architecture-based qualitative vulnerability assessment of representative
multi-agent AI pentesting frameworks according to the proposed attack taxonomy.
Attack categories: A=Pre-training Attacks, B=Fine-Tuning and Alignment Attacks,
C=Inference-Time Attacks, D=Memory and Knowledge Attacks,
E=Multi-Agent Coordination Attacks, F=Tool and Execution Attacks.}
\label{tab:multi_agent_vuln}

\setlength{\tabcolsep}{3pt}
\renewcommand{\arraystretch}{1.3}

\resizebox{\textwidth}{!}{%
\footnotesize
\begin{tabular}{@{}
>{\raggedright\arraybackslash}p{30mm}
>{\raggedright\arraybackslash}p{22mm}
>{\raggedright\arraybackslash}p{25mm}
*{6}{>{\centering\arraybackslash}m{6mm}}
>{\raggedright\arraybackslash}p{58mm}
@{}}

\toprule

\multirow{2}{*}{\textbf{Framework}}
&
\multirow{2}{*}{\textbf{Base LLM}}
&
\multirow{2}{*}{\textbf{Architecture}}
&
\multicolumn{6}{c}{\textbf{Attack Category}}
&
\multirow{2}{*}{\makecell[c]{\textbf{Architectural}\\\textbf{Security Observations}}}

\\

\cmidrule(lr){4-9}

&
&
&
\textbf{A}
&
\textbf{B}
&
\textbf{C}
&
\textbf{D}
&
\textbf{E}
&
\textbf{F}
&

\\

\midrule

\textbf{VulnBot} \cite{kong2025vulnbot}
&
GPT-4o / Llama-3
&
Penetration Testing Graph(PTG) + Retrieval-Augmented Generation(RAG) + Multi-Agent.
&
\Med
&
\Med
&
\High
&
\High
&
\High
&
\High
&
Persistent memory and PTG coordination increase exposure to memory,
coordination, and execution attacks.
\\

\textbf{BreachSeek} \cite{alshehri2024breachseek}
&
GPT-4o / Claude 3.5
&
Supervisor + Specialist Agents
&
\Med
&
\Med
&
\High
&
\Med
&
\High
&
\High
&
Shared task context and orchestrator communication create propagation
paths for coordination attacks.
\\

\textbf{PentestAgent} \cite{shen2025pentestagent}
&
GPT-4
&
Planning Agents + RAG + Execution History
&
\Med
&
\Med
&
\High
&
\High
&
\High
&
\High
&
Knowledge retrieval and inter-agent communication expose both memory and
coordination attack surfaces.
\\

\textbf{Incalmo} \cite{singer2025incalmo}
&
LLM-agnostic
&
Planner + Attack Graph + Specialist Agents.
&
\Low
&
\Low
&
\Med
&
\Med
&
\Med
&
\Med
&
Intent abstraction reduces direct prompt attacks, while shared attack
graphs remain susceptible to manipulation.
\\

\textbf{ReaperAI} \cite{valencia2024artificial}
&
GPT-4
&
Autonomous Multi-Phase.
&
\Med
&
\Med
&
\High
&
\Med
&
\High
&
\High
&
Autonomous execution and phase coordination significantly increase
execution-layer attack exposure.
\\

\textbf{CAI} \cite{mayoral2025cai}
&
Multi-LLM
&
Multi-Agent + Model Context Protocol (MCP).
&
\High
&
\High
&
\High
&
\High
&
\High
&
\High
&
Supports numerous foundation models and external MCP tools, resulting in
the broadest architectural attack surface among surveyed frameworks.
\\

\textbf{AutoPentest} \cite{henke2025autopentest}
&
GPT-4 / GPT-4o
&
Penetration Testing Tree (PTT) + Autonomous Multi-Agent.
&
\Med
&
\Med
&
\High
&
\Med
&
\High
&
\High
&
Autonomous planning and distributed execution increase exposure to
coordination failures and unsafe tool invocation.
\\

\textbf{PENTEST-AI} \cite{bianou2024pentest}
&
GPT-4
&
MITRE ATT\&CK-Guided Multi-Agent.
&
\Med
&
\Med
&
\High
&
\High
&
\High
&
\High
&
Multiple collaborative agents guided by MITRE ATT\&CK taxonomy expand 
the attack surface through shared reasoning, memory exchange, 
and autonomous execution.
\\

\textbf{AutoPentester} \cite{ginige2025autopentester}
&
GPT-4 / GPT-4o
&
Planner + Supervisor + Specialized Workers + Retrieval-Augmented Generation (RAG).
&
\Med
&
\Med
&
\High
&
\High
&
\High
&
\High
&
Supervisor-coordinated workers and RAG-based strategy retrieval increase
exposure to memory, coordination, and execution attacks; low subtask
completion reported due to strategy identification failures.
\\

\textbf{PTFusion} \cite{wang2025ptfusion}
&
GPT-4
&
MasterAgent + ReconAgents.
&
\Med
&
\Med
&
\High
&
\High
&
\High
&
\High
&
Distributed reconnaissance and shared memory improve coverage but expose
the framework to coordination and memory poisoning attacks.
\\

\textbf{xOffense} \cite{luong2025xoffense}
&
Multi-LLM
&
Collaborative Multi-Agent.
&
\High
&
\High
&
\High
&
\High
&
\High
&
\High
&
Heterogeneous multi-agent collaboration and extensive tool integration
create a broad attack surface across all taxonomy categories.
\\
\bottomrule

\end{tabular}

}

\vspace{2mm}

\footnotesize
\textit{Note:}
Ratings indicate qualitative architectural exposure derived from the
published framework designs and should not be interpreted as empirical
attack success rates.

\vspace{0.5mm}
\footnotesize
\textbf{Legend:}
\colorbox{red!25}{\textbf{H}}~High architectural exposure;
\colorbox{orange!30}{\textbf{M}}~Moderate architectural exposure;
\colorbox{green!20}{\textbf{L}}~Low architectural exposure;
\colorbox{gray!15}{--}~Attack category structurally not applicable.
\end{table*}
\section{Guardrail Mechanisms for AI Pentesting Agents}
\label{sec:guardrail}
The guardrail taxonomy is intentionally organized to mirror the attack taxonomy, enabling systematic mapping between attack surfaces and defense mechanisms rather than representing an independent classification.
\begin{figure*}[t]
\centering
\begin{adjustbox}{max width=\textwidth}
\begin{tikzpicture}[
  font=\small,
  >=Latex,
  titlenode/.style={
    font=\small\bfseries, align=center
  },
  axisbox/.style={
    draw, rounded corners=6pt, align=center,
    minimum height=10mm, text width=39mm,
    font=\small\bfseries, thick
  },
  guardrailbox/.style={
    draw, rounded corners=4pt, align=center,
    minimum height=9mm, text width=38mm,
    font=\small
  },
  attacklabel/.style={
    draw=none, align=center, font=\scriptsize\itshape,
    text width=36mm
  },
  connector/.style={draw, thick, -Latex},
  brace/.style={decorate,
    decoration={brace, amplitude=5pt, raise=2pt}}
]

\node[titlenode] at (-4.8, 0.3)  {\textbf{Guardrail}};
\node[titlenode] at (-4.8, -0.1) {\textbf{Category}};
\node[titlenode] at (0,    0.3)  {\textbf{Protected}};
\node[titlenode] at (0,   -0.1)  {\textbf{Attack Surface}};
\node[titlenode] at (4.8,  0.3)  {\textbf{Mitigated}};
\node[titlenode] at (4.8, -0.1)  {\textbf{Attack Cat.}};

\draw[thick, gray!60] (-7.2, -0.4) -- (7.2, -0.4);

\node[font=\small\bfseries, rotate=90, text=red!70!black]
  at (-7.0, -2.2) {LLM Lifecycle Layer};
\draw[thick, red!40, dashed]
  (-6.6, -0.5) -- (-6.6, -4.9);

\node[axisbox, fill=red!12, draw=red!60] (GA) at (-4.8, -1.1)
  {G-A\\Pre-training\\Guardrails};
\node[guardrailbox, fill=red!5, draw=red!40] (GAd) at (0, -1.1)
  {Dataset provenance,\\backdoor detection,\\frequency-space cleansing};
\node[attacklabel] (GAa) at (4.8, -1.1)
  {Cat.\,A\\Pre-Training \&\\Corpus Poisoning};

\node[axisbox, fill=red!12, draw=red!60] (GB) at (-4.8, -2.7)
  {G-B\\Fine-Tuning \&\\Alignment Guardrails};
\node[guardrailbox, fill=red!5, draw=red!40] (GBd) at (0, -2.7)
  {Tamper-resistant tuning,\\safe LoRA, representation\\noise, vaccine perturbation};
\node[attacklabel] (GBa) at (4.8, -2.7)
  {Cat.\,B\\Fine-Tuning \&\\Alignment Attacks};

\node[axisbox, fill=red!12, draw=red!60] (GC) at (-4.8, -4.3)
  {G-C\\Inference-Time\\Guardrails};
\node[guardrailbox, fill=red!5, draw=red!40] (GCd) at (0, -4.3)
  {Input/output classifiers,\\injection detection,\\constitutional AI};
\node[attacklabel] (GCa) at (4.8, -4.3)
  {Cat.\,C\\Inference-Time\\Prompt Attacks};

\draw[brace, red!50]
  (-6.5, -0.6) -- (-6.5, -4.9)
  node[midway, left=6pt, font=\scriptsize\bfseries,
       text=red!70!black, align=center]
  {};

\draw[thick, gray!40, dashed] (-6.6, -4.9) -- (7.0, -4.9);

\node[font=\small\bfseries, rotate=90, text=blue!70!black]
  at (-7.0, -7.2) {Agent Architecture Layer};

\node[axisbox, fill=blue!10, draw=blue!50] (GD) at (-4.8, -5.6)
  {G-D\\Memory \&\\Knowledge Guardrails};
\node[guardrailbox, fill=blue!5, draw=blue!40] (GDd) at (0, -5.6)
  {Provenance tracking,\\consensus validation,\\activation-based detection};
\node[attacklabel] (GDa) at (4.8, -5.6)
  {Cat.\,D\\Memory \&\\Knowledge Attacks};

\node[axisbox, fill=blue!10, draw=blue!50] (GE) at (-4.8, -7.2)
  {G-E\\Multi-Agent\\Coordination Guardrails};
\node[guardrailbox, fill=blue!5, draw=blue!40] (GEd) at (0, -7.2)
  {Trust verification,\\message authentication,\\worm containment};
\node[attacklabel] (GEa) at (4.8, -7.2)
  {Cat.\,E\\Multi-Agent\\Coordination Attacks};

\node[axisbox, fill=blue!10, draw=blue!50] (GF) at (-4.8, -8.8)
  {G-F\\Tool \&\\Execution Guardrails};
\node[guardrailbox, fill=blue!5, draw=blue!40] (GFd) at (0, -8.8)
  {Tool-call verification,\\temporal constraints,\\plugin/MCP validation};
\node[attacklabel] (GFa) at (4.8, -8.8)
  {Cat.\,F\\Tool \&\\Execution Attacks};

\draw[brace, blue!50]
  (-6.5, -5.0) -- (-6.5, -9.3)
  node[midway, left=6pt, font=\scriptsize\bfseries,
       text=blue!70!black, align=center]
  {};

\foreach \src/\mid/\dst in {
  GA/GAd/GAa, GB/GBd/GBa, GC/GCd/GCa,
  GD/GDd/GDa, GE/GEd/GEa, GF/GFd/GFa}
{
  \draw[connector] (\src.east) -- (\mid.west);
  \draw[connector] (\mid.east) -- (\dst.west);
}

\node[font=\scriptsize\itshape, align=center, text=gray!70]
at (0,-9.9)
{G-A through G-C mitigate vulnerabilities inherited from the
foundation model lifecycle, whereas G-D through G-F protect
architectural components introduced by autonomous AI agent
frameworks.};

\end{tikzpicture}
\end{adjustbox}
\caption{Proposed guardrail taxonomy for autonomous AI pentesting agents,
organized along the same two-axis structure as the attack taxonomy.
Categories G-A--G-C mitigate vulnerabilities inherited from the LLM
lifecycle, whereas Categories G-D--G-F protect the architectural
components introduced by autonomous agent frameworks. Each guardrail
category is designed to defend the primary attack surface associated
with its corresponding attack category, establishing a one-to-one
mapping between the proposed attack and defense taxonomies.}
\label{fig:guardrail_taxonomy}
\end{figure*}

While substantial progress has been made in developing guardrail mechanisms
for large language models (LLMs), most existing approaches have been designed
for conversational assistants, content moderation, or general-purpose AI
applications. Autonomous AI pentesting agents operate in a fundamentally
different environment, where they must reason over untrusted observations,
maintain long-term memory, coordinate with other agents, and autonomously
interact with external security tools. Consequently, guardrails designed for
conventional LLMs are insufficient to address the broader attack surface
introduced by agentic AI systems.

To systematically organize existing defense mechanisms, this survey proposes
a two-axis guardrail taxonomy that mirrors the attack taxonomy presented in
Section~\ref{sec:taxonomy}. The first axis, \emph{LLM Lifecycle Guardrails}
(Categories G-A--G-C), protects the foundation model throughout its
development lifecycle, including pre-training, fine-tuning and alignment, and
inference. The second axis, \emph{Agent Architecture Guardrails}
(Categories G-D--G-F), protects the architectural components introduced by
autonomous AI agents, including persistent memory, retrieval mechanisms,
multi-agent coordination, and external tool execution.

Unlike conventional LLM guardrails that primarily focus on preventing unsafe
text generation, the proposed taxonomy extends protection across multiple
architectural trust boundaries. Accordingly, each guardrail category is
designed to mitigate the primary attack surface identified in the
corresponding attack taxonomy, thereby establishing a one-to-one mapping
between attacks and defenses.

Figure~\ref{fig:guardrail_taxonomy} illustrates the proposed guardrail
taxonomy. Categories G-A--G-C mitigate vulnerabilities inherited from the
foundation model lifecycle, whereas Categories G-D--G-F protect the
agent-specific architectural components that enable autonomous penetration
testing. Together, these guardrail categories provide a comprehensive defense
framework for securing AI-based pentesting agents against both inherited LLM
vulnerabilities and emergent agent-specific attacks.


\subsection{G-A: Pre-Training Guardrails}
\label{subsec:ga}

Pre-training guardrails protect the integrity of the foundation model by
securing the large-scale corpus and optimization pipeline used during
pre-training. Since vulnerabilities introduced at this stage propagate to
all downstream applications, these guardrails aim to detect and eliminate
poisoned data, hidden backdoors, and malicious training samples before the
foundation model is deployed.

\textbf{Protected Component.}
The primary protected components include the pre-training corpus, data
collection pipeline, and model optimization process.

\textbf{Protected Trust Boundary.}
G-A guardrails protect the \emph{Model Boundary} by ensuring that the
foundation model is trained using trustworthy data prior to deployment.

\textbf{Representative Defense Mechanisms.}
Representative approaches include dataset provenance verification (Gracefully)~\cite{wu2025gracefully} that identifies poisoned samples through importance-weighted resampling, activation-space backdoor detection~\cite{li2026backdoorllm} that detects hidden trigger behaviours by analysing internal model representations, and chain-of-thought based backdoor detection~\cite{li2025chain} that exploits reasoning discontinuities exhibited by compromised models. The combination of these techniques results in the improved integrity of the pre-training pipeline by identifying malicious training data before model optimisation. Another line of work focuses on improving clean-model recovery after poisoning by downscaling frequency-space artefacts in backdoored training data~\cite{wu2024acquiring}, which provides a corpus-cleansing mechanism different than provenance-based filtering.

\subsection{G-B: Fine-Tuning and Alignment Guardrails}
\label{subsec:gb}

Fine-tuning and alignment guardrails protect the post-training adaptation
pipeline by ensuring that supervised fine-tuning (SFT), reinforcement
learning from human feedback (RLHF), direct preference optimization (DPO),
and parameter-efficient fine-tuning methods preserve the safety and
alignment properties of the foundation model. These guardrails seek to
prevent adversarial manipulation of the model during specialization while
maintaining its intended behavior.

\textbf{Protected Component.}
The primary protected components include fine-tuning datasets, preference
annotations, reward models, alignment objectives, and parameter-efficient
adaptation modules such as LoRA adapters.

\textbf{Protected Trust Boundary.}
G-B guardrails protect the \emph{Model Boundary} by securing the
post-training adaptation pipeline before the model is deployed.

\textbf{Representative Defense Mechanisms.}
Representative approaches include tamper-resistant safety
fine-tuning~\cite{tamirisa2025tamper}, which improves the robustness of
aligned models against subsequent malicious fine-tuning, Safe
LoRA~\cite{hsu2024safe}, which constrains adapter updates within a
safety-preserving subspace, PEFTGuard~\cite{sun2025peftguard}, which audits
parameter-efficient adapters for malicious modifications, representation
noising~\cite{rosati2024representation}, which reduces the effectiveness of
backdoor triggers through controlled perturbations of latent
representations, Vaccine~\cite{huang2024vaccine}, which pre-immunizes
models against alignment attacks using simulated adversarial exposure, and
SEAL~\cite{shen2025seal}, which improves alignment robustness through
careful selection of fine-tuning data. Further defenses in this space
include Antidote~\cite{huang2024antidote} and Booster~\cite{huang2025booster},
which counteract harmful fine-tuning by attenuating malicious gradient
perturbations post-hoc; SaLoRA~\cite{li2025salora}, which preserves safety
alignment within low-rank adaptation updates; P2P~\cite{zhao2025p2p}, a
poison-to-poison remedy for backdoor defense; and SCOUT~\cite{afane2025scout},
which detects data-poisoning attempts during fine-tuning.

\subsection{G-C: Inference-Time Guardrails}
\label{subsec:gc}

Inference-time guardrails protect AI systems during deployment by
monitoring and controlling interactions between users, the LLM, and the
external environment. Unlike pre-training and alignment defenses, these
mechanisms operate at runtime to detect prompt injection, jailbreak
attempts, malicious tool outputs, and unsafe model responses before they
can influence the agent's reasoning or execution pipeline.

\textbf{Protected Component.}
The core protected components are user prompts, system prompts, reasoning context, model responses and runtime interactions between the LLM and external environments.

\textbf{Protected Trust Boundary.}
G-C guardrails protect the \emph{Context Boundary} by making sure that untrusted runtime inputs do not influence a model's reasoning process or compromise its decision making during deployment.

\textbf{Representative Defense Mechanisms.}
Inference-time defences approach the problem from multiple angles. LlamaGuard~\cite{inan2023llama} performs prompt and response classification for safety. PromptShield~\cite{jacob2024promptshield} aims at direct and indirect prompt injection . Constitutional AI~\cite{bai2022constitutional} instead limits behaviour through principle-based self-criticism. Framework-level tools like NeMo Guardrails~\cite{rebedea2023nemo} enforce policies on conversation and tool execution, while Guardrails AI~\cite{dong2024building} validates inputs and outputs against user-defined schemas. DataSentinel~\cite{liu2025datasentinel} monitors runtime inputs for adversarial patterns. Other works include design patterns for securing LLM agents against injection~\cite{beurer2025design}, Attention Tracker~\cite{hung2025attention} that detects unusual attention patterns, SecurityLingua~\cite{li2025securitylingua} that compresses the prompts to remove malicious content, and AdaptiveGuard~\cite{yang2025adaptiveguard} that adapts runtime safety policies. A recent study offers a more comprehensive assessment of the effectiveness of jailbreak guardrails~\cite{wang2025sok} while another suggests a proactive defence against jailbreak~\cite{zhao2025proactive}. These mechanisms work together to improve deployed LLMs against inference-time manipulation without affecting normal functionality. Recent work on reasoning-heavy models further demonstrates that long chain-of-thought generation introduces safety failure modes that standard input/output classifiers miss \cite{jiang2025safechain}.

\subsection{G-D: Memory and Knowledge Guardrails}
\label{subsec:gd}

Memory and knowledge guardrails preserve the integrity of external knowledge sources that augment the reasoning power of autonomous AI agents. Unlike the static parameters of the foundation model, these knowledge repositories are updated dynamically during deployment and may include retrieval-augmented generation (RAG) systems, vector databases, episodic memory, long-term memory, knowledge graphs, and external document repositories. Therefore, these guardrails aim to ensure that retrieved information is reliable before it influences the agent’s reasoning process. 

\textbf{Protected Component.}
The main protected components are vector databases, retrieval indices, external knowledge repositories, episodic memory, long-term memory, and other persistent knowledge stores used by autonomous AI agents. 

\textbf{Protected Trust Boundary.}
G-D guardrails protect the \emph{Context Boundary} by preventing malicious manipulation of the retrieved knowledge and stored memory before they are integrated into the agent's reasoning process.

\textbf{Representatives Defense Mechanisms.}
Representative approaches include TrustRAG~\cite{zhou2025trustrag} which validates retrieved documents via trust-aware retrieval, A-MemGuard~\cite{wei2025memguard} which detects memory poisoning by monitoring abnormal memory updates, MAGE~\cite{wang2026mage} which evaluates the reliability of retrieved evidence before reasoning, RevPRAG~\cite{tan2024revprag} which improves retrieval robustness against poisoned knowledge sources through evidence verification, and activation-based memory integrity verification techniques which identify malicious knowledge before it is incorporated into the agent’s internal reasoning state. Complementary work includes certifiably robust RAG against retrieval corruption~\cite{xiang2024certifiably}, RAGuard~\cite{kolhe2025raguard} that presents a layered defence framework against RAG data poisoning, and work on tracing and attributing poisoned knowledge back to its source within RAG pipelines~\cite{zhang2025taught,zhang2025traceback}. Together, these mechanisms improve the robustness of retrieval and persistent memory while alleviating the threat of knowledge poisoning attacks in autonomous AI agents.

\subsection{G-E: Multi-Agent Coordination Guardrails}
\label{subsec:ge}

Multi-agent coordination guardrails protect the communication and coordination mechanisms that enable multiple autonomous agents to cooperate toward a common objective. As AI pentesting frameworks evolve toward multi-agent architectures, agents communicate through shared channels, exchanging plans, observations, retrieved knowledge and execution results. These guardrails are aimed at preserving the integrity, authenticity and trustworthiness of inter-agent interactions and hence avoid compromised agents from affecting the collective decision-making process.

\textbf{Protected Component.}
Important protected elements are communication channels between the agents, shared task representations, orchestration frameworks, coordination protocols, and trust relations between collaborating agents.

\textbf{Protected Trust Boundary.}
G-E guardrails protect the \emph{Execution Boundary} by verifying the authenticity, trustworthiness and resistance to manipulation of the messages exchanged between cooperating agents during the execution lifecycle.

\textbf{Representative Defense Mechanisms.}
Representative approaches include cryptographic message authentication for verifying inter-agent communication, trust-aware agent verification frameworks that dynamically evaluate the reliability of participating agents, consensus-based coordination mechanisms that validate shared decisions before execution, and RouteGuard~\cite{xiao2026routeguard} that secures multi-agent communication paths by detecting malicious routing behaviours. Other techniques include verification protocols for communication, role-based authorisation mechanisms, and crossagent consistency checking that detect conflicting observations or abnormal patterns of coordination before they are propagated throughout the network of agents. Further approaches include AutoDefense~\cite{zeng2024autodefense}, which coordinates multiple LLM agents to jointly filter jailbreak attempts; a multi-agent defense pipeline against prompt injection~\cite{hossain2025multi}; CoopGuard~\cite{li2026coopguard}, which safeguards cooperative agents against evolving multi-round attacks; TrinityGuard~\cite{wang2026trinityguard}, a unified framework for securing multi-agent systems; PSG-Agent~\cite{wu2025psg}, a personality-aware safety guardrail for LLM-based agents; and work on reconstructing cross-agent semantic flows for execution-aware attack detection~\cite{wei2026beyond}. These mechanisms together increase the robustness of multi-agent systems against compromised agents and malicious message propagation with collaborative reasoning.

\subsection{G-F: Tool and Execution Guardrails}
\label{subsec:gf}

Tool and execution guardrails protect the interfaces that autonomous AI agents use to interact with external environments. As opposed to typical LLM use cases where the model just generates text responses, AI pentesting agents run shell commands, call security tools, query external services, and coordinate with plugins via standardised interfaces like function calling and the Model Context Protocol (MCP). These guardrails confirm the generated actions before execution, reducing the risk that the actions are unsafe, unintended, or adversarially manipulated.

\textbf{Protected Component.}
The primary protected elements include tool invocation interfaces, APIs for function-calling, modules for shell execution, plugin ecosystems, MCP servers, execution workflows, and integrations with external services.

\textbf{Protected Trust Boundary.}
G-F guardrails secure the \emph{Execution Boundary} by requiring validation of actions generated by LLMs before interfacing with external systems or security tools. 

\textbf{Representative Defense Mechanisms.}
Representative approaches include ToolSafe~\cite{mou2026toolsafe}, which enforces proactive step-level guardrails and feedback before a tool call is allowed to execute; TraceSafe~\cite{chen2026tracesafe}, which evaluates guardrail effectiveness across full multi-step tool-calling trajectories rather than individual calls in isolation; work on enforcing temporal constraints for LLM agents~\cite{kamath2025enforcing}, which restricts the ordering and timing of tool invocations to prevent unsafe execution sequences; MindGuard~\cite{wang2025mindguard}, which inspects an agent's internal decision process to detect metadata poisoning before a tool is invoked; and SafeAgent~\cite{liu2026safeagent}, which provides a runtime protection layer that mediates and validates tool and plugin execution for agentic systems.  These mechanisms, working together, reduce the likelihood of unauthorised tools being run, workflows being tampered with, the supply-chain being compromised, and malicious plugins being exploited, while retaining the necessary operational capabilities for autonomous penetration testing.

The qualitative assessment in Tables~\ref{tab:guardrail_single} and
\ref{tab:guardrail_multi} is derived from the architectural design and
guardrail mechanisms explicitly documented in each framework's
publication, rather than empirical attack evaluation. Each entry
identifies the dominant architectural factor limiting guardrail
effectiveness in that category; ``---'' denotes structural
inapplicability.
\begin{table*}[t]
\centering
\caption{Assessment of guardrail insufficiencies in representative
single-agent AI pentesting frameworks. G-A--G-C: LLM Lifecycle
Guardrails; G-D--G-F: Agent Architecture Guardrails.}
\label{tab:guardrail_single}
\scriptsize
\setlength{\tabcolsep}{4pt}
\renewcommand{\arraystretch}{1.3}
\resizebox{\textwidth}{!}{%
\begin{tabular}{
>{\raggedright\arraybackslash}p{25mm}
>{\raggedright\arraybackslash}p{26mm}
*{6}{>{\centering\arraybackslash}m{7mm}}
>{\raggedright\arraybackslash}p{37mm}}
\toprule
\multirow{2}{*}{\textbf{Tool (Year)}} &
\multirow{2}{*}{\textbf{Architecture}} &
\multicolumn{3}{c}{\textbf{LLM Lifecycle}} &
\multicolumn{3}{c}{\textbf{Agent Architecture}} &
\multirow{2}{*}{\textbf{Research Gap}} \\
\cmidrule(lr){3-5}\cmidrule(lr){6-8}
& & \textbf{G-A} & \textbf{G-B} & \textbf{G-C} & \textbf{G-D} & \textbf{G-E} & \textbf{G-F} & \\
\midrule

PentestGPT (2024)~\cite{deng2024pentestgpt} & Parsing–Reasoning–Generation &
\cellcolor{red!20}FM & \cellcolor{red!20}FM & \cellcolor{red!20}\makecell{OFN\\ENV} &
\cellcolor{orange!25}UNT & \cellcolor{gray!15}--- & \cellcolor{orange!25}SEM &
Human in the Loop reduces unsafe execution; memory and tool-call authorization remain weak. \\

AutoAttacker (2024)~\cite{xu2024autoattacker} & Planner + Retrieval-Augmented Generation (RAG) &
\cellcolor{red!20}FM & \cellcolor{red!20}FM & \cellcolor{red!20}\makecell{OFN\\ENV} &
\cellcolor{red!20}UNT & \cellcolor{gray!15}--- & \cellcolor{red!20}SEM &
Unverified RAG memory and unrestricted tool invocation dominate. \\

ARACNE (2025)~\cite{nieponice2025aracne} & Planner + Interpreter + Tool Execution &
\cellcolor{red!20}FM & \cellcolor{red!20}FM & \cellcolor{orange!25}OFN &
\cellcolor{gray!15}--- & \cellcolor{gray!15}--- & \cellcolor{orange!25}SEM &
Injection resistance improved; SSH execution lacks authorization control. \\

HackSynth (2024)~\cite{muzsai2024hacksynth} & Planner + Command Generation + Summarization &
\cellcolor{red!20}FM & \cellcolor{red!20}FM & \cellcolor{red!20}\makecell{OFN\\ENV} &
\cellcolor{gray!15}--- & \cellcolor{gray!15}--- & \cellcolor{red!20}SEM &
Autonomous shell execution without verification is the primary concern. \\

PenHeal (2023)~\cite{huang2023penheal} & Pentest + Remediation &
\cellcolor{red!20}FM & \cellcolor{red!20}FM & \cellcolor{orange!25}OFN &
\cellcolor{gray!15}--- & \cellcolor{gray!15}--- & \cellcolor{orange!25}SEM &
Partial validation cannot fully prevent adversarial influence on remediation. \\

HackingBuddyGPT (2023)~\cite{happe2023getting} & SSH Executor + Feedback &
\cellcolor{red!20}FM & \cellcolor{red!20}FM & \cellcolor{red!20}\makecell{OFN\\ENV} &
\cellcolor{gray!15}--- & \cellcolor{gray!15}--- & \cellcolor{red!20}SEM &
Direct shell execution without validation yields the highest exposure. \\

CHECKMATE (2025)~\cite{wang2025automated} & Planner + Validator &
\cellcolor{red!20}FM & \cellcolor{red!20}FM & \cellcolor{orange!25}OFN &
\cellcolor{gray!15}--- & \cellcolor{gray!15}--- & \cellcolor{orange!25}SEM &
Validation-guided reasoning improves robustness, but execution authorization remains limited. \\

Cochise (2026)~\cite{happe2026cochise} & Planner + Tool Execution + Benchmark Evaluation &
\cellcolor{red!20}FM & \cellcolor{red!20}FM & \cellcolor{orange!25}OFN &
\cellcolor{gray!15}--- & \cellcolor{gray!15}--- & \cellcolor{orange!25}SEM &
Controlled evaluation reduces attack exposure, although tool authorization is not explicitly enforced. \\

RefPentester (2025)~\cite{dai2025refpentester} & Planner + Reflection &
\cellcolor{red!20}FM & \cellcolor{red!20}FM & \cellcolor{orange!25}OFN &
\cellcolor{gray!15}--- & \cellcolor{gray!15}--- & \cellcolor{orange!25}SEM &
Self-reflection improves planning reliability, but prompt injection and unsafe tool execution remain possible. \\

\bottomrule
\end{tabular}
}

\vspace{0.5mm}
\scriptsize
\textbf{Legend:}
\textbf{FM}=Inherited protection from proprietary foundation models (guardrails are externally defined and cannot be independently verified or customized);
\textbf{OFN}=Conflict between offensive penetration-testing objectives and foundation-model safety filtering;
\textbf{ENV}=Adversarial observations accumulated during interaction that bias subsequent reasoning and planning;
\textbf{UNT}=Untrusted memory or knowledge provenance without integrity verification;
\textbf{SEM}=Semantic ambiguity in generated commands prior to execution, increasing the risk of unsafe tool invocation.

\vspace{0.5mm}
\noindent\scriptsize
\textit{Note:} G-A and G-B reflect protection inherited from the underlying foundation model rather than framework-level guardrails. G-E is structurally inapplicable to single-agent systems, as no inter-agent communication channel exists.

\vspace{0.5mm}
\noindent\scriptsize
\textbf{Color Coding:}\quad
\colorbox{red!20}{\hspace{10pt}} High Guardrail Insufficiency \quad
\colorbox{orange!25}{\hspace{10pt}} Moderate Guardrail Insufficiency \quad
\colorbox{green!20}{\hspace{10pt}} Low Guardrail Insufficiency \quad
\colorbox{gray!15}{\hspace{10pt}} Not Applicable
\end{table*}


\begin{table*}[t]
\centering
\caption{Assessment of guardrail insufficiencies in representative
multi-agent AI pentesting frameworks. G-A--G-C: LLM Lifecycle
Guardrails; G-D--G-F: Agent Architecture Guardrails.}
\label{tab:guardrail_multi}
\scriptsize
\setlength{\tabcolsep}{4pt}
\renewcommand{\arraystretch}{1.3}
\resizebox{\textwidth}{!}{%
\begin{tabular}{
>{\raggedright\arraybackslash}p{25mm}
>{\raggedright\arraybackslash}p{26mm}
*{6}{>{\centering\arraybackslash}m{7mm}}
>{\raggedright\arraybackslash}p{37mm}}
\toprule
\multirow{2}{*}{\textbf{Tool (Year)}} &
\multirow{2}{*}{\textbf{Architecture}} &
\multicolumn{3}{c}{\textbf{LLM Lifecycle}} &
\multicolumn{3}{c}{\textbf{Agent Architecture}} &
\multirow{2}{*}{\textbf{Research Gap}} \\
\cmidrule(lr){3-5}\cmidrule(lr){6-8}
& & \textbf{G-A} & \textbf{G-B} & \textbf{G-C} & \textbf{G-D} & \textbf{G-E} & \textbf{G-F} & \\
\midrule

VulnBot (2025)~\cite{kong2025vulnbot} & Penetration Testing Graph (PTG) + Retrieval-Augmented Generation (RAG) + Multi-Agent &
\cellcolor{red!20}FM & \cellcolor{red!20}FM & \cellcolor{red!20}OFN &
\cellcolor{red!20}UNT & \cellcolor{red!20}ROL & \cellcolor{red!20}SEM &
Unverified task-graph coordination and RAG memory enable cross-agent propagation. \\

BreachSeek (2024)~\cite{alshehri2024breachseek} & Supervisor + Specialist Agents &
\cellcolor{red!20}FM & \cellcolor{red!20}FM & \cellcolor{red!20}OFN &
\cellcolor{orange!25}UNT & \cellcolor{red!20}ROL & \cellcolor{red!20}SEM &
Absence of trust verification enables adversarial propagation across specialists. \\

PentestAgent (2024)~\cite{shen2025pentestagent} & Reconnaissance Agent + Search Agents + Retrieval-Augmented Generation (RAG) &
\cellcolor{red!20}FM & \cellcolor{red!20}FM & \cellcolor{red!20}OFN &
\cellcolor{red!20}UNT & \cellcolor{red!20}ROL & \cellcolor{red!20}SEM &
Untrusted retrieval and unrestricted messaging remain dominant weaknesses. \\

Incalmo (2025)~\cite{singer2025incalmo} & Planner + Attack Graph + Specialist Agents &
\cellcolor{red!20}FM & \cellcolor{red!20}FM & \cellcolor{orange!25}OFN &
\cellcolor{orange!25}UNT & \cellcolor{orange!25}ROL & \cellcolor{orange!25}SEM &
Modular isolation reduces propagation; cross-stage trust verification remains limited. \\

ReaperAI (2024)~\cite{valencia2024artificial} & Autonomous Multi-Phase &
\cellcolor{red!20}FM & \cellcolor{red!20}FM & \cellcolor{red!20}OFN &
\cellcolor{orange!25}UNT & \cellcolor{red!20}ROL & \cellcolor{red!20}SEM &
Full autonomy lacks runtime verification for coordination and execution. \\

CAI (2025)~\cite{mayoral2025cai} & Multi-Agent + Model Context Protocol (MCP) &
\cellcolor{red!20}FM & \cellcolor{red!20}FM & \cellcolor{red!20}OFN &
\cellcolor{red!20}UNT & \cellcolor{red!20}ROL & \cellcolor{red!20}SEM &
Model diversity and unrestricted MCP/plugin execution yield the broadest attack surface. \\

AutoPentest (2024)~\cite{henke2025autopentest} & Penetration Testing Tree (PTT) + Autonomous Multi-Agent &
\cellcolor{red!20}FM & \cellcolor{red!20}FM & \cellcolor{red!20}OFN &
\cellcolor{orange!25}UNT & \cellcolor{red!20}ROL & \cellcolor{red!20}SEM &
Autonomous agent collaboration lacks runtime trust verification and secure tool authorization. \\

PENTEST-AI (2025)~\cite{bianou2024pentest} & MITRE ATT\&CK-Guided Multi-Agent. &
\cellcolor{red!20}FM & \cellcolor{red!20}FM & \cellcolor{red!20}OFN &
\cellcolor{red!20}UNT & \cellcolor{red!20}ROL & \cellcolor{red!20}SEM &
Collaborative reasoning lacks authenticated coordination and secure memory validation mechanisms. \\

AutoPentester (2025)~\cite{ginige2025autopentester} & Planner + Supervisor + Specialized Workers + RAG. &
\cellcolor{red!20}FM & \cellcolor{red!20}FM & \cellcolor{red!20}OFN &
\cellcolor{red!20}UNT & \cellcolor{red!20}ROL & \cellcolor{red!20}SEM &
Supervisor-mediated task delegation and RAG retrieval lack authenticated coordination and provenance verification. \\

PTFusion (2025)~\cite{wang2025ptfusion} & MasterAgent + ReconAgents &
\cellcolor{red!20}FM & \cellcolor{red!20}FM & \cellcolor{red!20}OFN &
\cellcolor{red!20}UNT & \cellcolor{orange!25}ROL & \cellcolor{red!20}SEM &
Distributed reconnaissance increases shared-memory exposure while inter-agent trust verification remains incomplete. \\

xOffense (2025)~\cite{luong2025xoffense} & Collaborative Multi-Agent &
\cellcolor{red!20}FM & \cellcolor{red!20}FM & \cellcolor{red!20}OFN &
\cellcolor{red!20}UNT & \cellcolor{red!20}ROL & \cellcolor{red!20}SEM &
Extensive multi-agent collaboration and heterogeneous tool integration lack unified guardrail enforcement. \\

\bottomrule
\end{tabular}
}

\vspace{0.5mm}
\scriptsize
\textbf{Legend:}
\textbf{FM}=Inherited protection from proprietary foundation models (guardrails are externally defined and cannot be independently verified or customized);
\textbf{OFN}=Conflict between offensive penetration-testing objectives and foundation-model safety filtering;
\textbf{ENV}=Adversarial observations accumulated during interaction that bias subsequent reasoning and planning;
\textbf{UNT}=Untrusted memory or knowledge provenance without integrity verification;
\textbf{ROL}=Absence of role-based trust verification between cooperating agents, allowing a compromised agent to influence others without authentication;
\textbf{SEM}=Semantic ambiguity in generated commands prior to execution, increasing the risk of unsafe tool invocation.

\vspace{0.5mm}
\noindent\scriptsize
\textit{Note:} G-A and G-B reflect protection inherited from the underlying foundation model rather than framework-level guardrails. G-D--G-F assess framework-specific guardrails governing persistent memory, multi-agent coordination, and autonomous tool execution.

\vspace{1mm}

\scriptsize
\textbf{Color Coding:}\quad
\colorbox{red!20}{\hspace{10pt}} High Guardrail Insufficiency \quad
\colorbox{orange!25}{\hspace{10pt}} Moderate Guardrail Insufficiency \quad
\colorbox{green!20}{\hspace{10pt}} Low Guardrail Insufficiency \quad
\colorbox{gray!15}{\hspace{10pt}} Not Applicable
\end{table*}

\section{Research Gaps and Future Directions}
\label{sec:gaps}

As analysed in the sections above, existing guardrail mechanisms are disjoint across the layers of LLM lifecycle and agent architecture. Despite great progress in the mitigation of individual attack classes, current defences mostly operate in isolation and only partially protect autonomous AI pentesting agents. The results of the evaluation in Tables~\ref{tab:guardrail_single} and~\ref{tab:guardrail_multi} suggest some promising avenues for future research.

\textbf{RG1 --- Integrated architecture-aware guardrail frameworks.}
Existing guardrails generally focus on protecting individual components like training pipelines, inference prompts, memory stores or external tool interfaces. However, autonomous AI pentesting agents integrate all these
components within a single execution pipeline. Future research should
develop unified guardrail frameworks capable of simultaneously securing
the LLM lifecycle layer and the agent architecture layer while providing
end-to-end protection across multiple trust boundaries.

\textbf{RG2 --- Runtime verification of foundation model integrity.}
Although numerous techniques have been proposed for detecting
pre-training poisoning, alignment manipulation, and model backdoors,
these approaches are largely designed for offline model evaluation.
Little attention has been devoted to continuously verifying the
integrity of foundation models deployed within autonomous AI
pentesting agents. Runtime verification techniques capable of detecting
compromised model behaviour remain an important open challenge.

\textbf{RG3 --- Execution-aware safety reasoning.}
Current guardrail mechanisms predominantly analyze textual prompts and
responses while providing limited reasoning about the operational
semantics of autonomous tool execution. Future guardrails should
incorporate execution-aware reasoning capable of evaluating command
intent, execution context, authorization boundaries, and the potential
security impact of downstream actions before tool invocation.

\textbf{RG4 --- Trustworthy memory and knowledge management.}
Persistent memory, retrieval-augmented generation (RAG), and shared
knowledge repositories have become fundamental components of autonomous
AI pentesting agents. However, memory provenance verification, integrity
validation, continual consistency checking, and anomaly detection remain
relatively immature. An important direction of research is to design trustworthy memory management architectures that are robust against long-term poisoning attacks.

\textbf{RG5 --- Secure multi-agent coordination.}
With the growing use of multi-agent pentesting frameworks, the demand for reliable cooperation between autonomous agents is crucial. Existing systems offer little or no support for agent authentication, message integrity validation, trust establishment or collaborative decision validation. Further research should focus on safe communication protocols, trust-aware coordination methods, and defences against adversarial message passing and worm-like attacks in collaborative AI ecosystems.

\textbf{RG6 --- Secure execution and invocation control of tools}
AI pentesting agents are increasingly using external plugins, Model Context Protocol (MCP) servers, browser automation, and autonomous shell execution, all of which significantly increase the attack surface. Existing guardrails generally do not attempt to verify execution intent or enforce fine-grained authorisation policies prior to invoking external tools. We recommend the development of policy-aware execution frameworks that can validate tool requests and enforce operational constraints and prevent unauthorised or unintentional actions in future work. 

\textbf{RG7 --- Standardised benchmarks and evaluation methodologies.}
While existing evaluation frameworks such as AgentDojo~\cite{debenedetti2024agentdojo}, AgentLAB~\cite{jiang2026agentlab}, LITMUS~\cite{zhang2026litmus} and work measuring agent progress on multi-step cyber attack scenarios~\cite{folkerts2026measuring} capture important aspects of autonomous agents, they do not offer a complete evaluation of AI pentesting systems in light of the taxonomy we propose. Future work should create standardised benchmarks, attack suites, and quantitative metrics to evaluate the effectiveness of guardrails, attack containment, execution safety, and long-term system robustness in realistic adversarial settings. 

\textbf{RG8 --- Governance, auditability, and accountability.}
In addition to technical defences, the safe deployment of autonomous AI pentesting agents requires rigorous governance mechanisms.
Standardised techniques for execution auditing, decision provenance, memory traceability, authorisation logging, and accountability have been little explored. The development of such governance frameworks will be critical in enabling safe and transparent deployment of autonomous offensive artificial intelligence systems in the real world.

Frameworks like RedTeamLLM~\cite{challita2025redteamllm} and more general treatments of offensive-security ideas and practices for artificial intelligence~\cite{harguess2025offensive} show early efforts at operationalising agentic red-teaming, but do not yet include the architecture-aware guardrail coverage that this survey argues is necessary.

Overall, these research gaps indicate that the current guardrail mechanisms are still mostly component-specific and provide limited protection against cross-layer attacks on autonomous AI pentesting agents. Tackling these challenges will require integrated, architecture-aware defence frameworks that can simultaneously secure foundation models, persistent memory, multi-agent collaboration, and autonomous tool execution. Developing such holistic guardrail architectures is a key research direction toward trustworthy autonomous AI-based penetration testing systems.

\section{Conclusion}
\label{sec:conclusion}

The rapid evolution of large language models has revolutionised the field of AI-based penetration testing from interactive assistant systems to autonomous agents who can reason, plan, manage memory and execute tasks with the aid of tools. These capabilities significantly enhance the efficiency of penetration testing, but they also present a wide and dynamic attack surface beyond the traditional realm of LLM security.
Therefore, the protection of autonomous AI pentesting agents has to take into account defences that span both the foundation model lifecycle and the agent architecture covering vulnerabilities.

This survey presents a broad analysis of the security landscape of autonomous AI pentesting agents. Firstly, we reviewed typical frameworks for single-agent and multi-agent systems, outlining their architectural designs and operational characteristics. Based on this analysis, we proposed a two-axis attack taxonomy that distinguishes vulnerabilities introduced by the \emph{LLM Lifecycle Layer} (pre-training, fine-tuning, and inference-time attacks) from those introduced by the \emph{Agent Architecture Layer} (memory and knowledge poisoning, multi-agent coordination attacks, and tool and execution attacks). This taxonomy offers a unified view for systematically classifying not only inherited vulnerabilities of LLMs but also threats specific to the agent.

We applied this taxonomy to assess representative AI pentesting frameworks and showed that the recent systems, despite their diverse architectural design, remain largely exposed to several attack categories. We further proposed a complementary guardrail taxonomy and studied how well the attack surfaces are mitigated by existing defensive mechanisms. Our analysis reveals that existing guardrails are still isolated and mainly protect isolated components, providing little protection against attacks that propagate through multiple trust boundaries.

The research gaps discussed in this survey also suggest that future AI pentesting agents will need integrated, architecture-aware guardrail frameworks that can simultaneously protect foundation models, persistent memory, multi-agent collaboration, and autonomous tool execution. Furthermore, standardised evaluation benchmarks, reliable governance mechanisms, and runtime verification techniques will be necessary for the reliable deployment of autonomous offensive artificial intelligence systems.

In summary, this survey offers a unified framework for understanding the evolving attack landscape facing AI-based penetration testing agents, and makes the case that securing them demands holistic mechanisms addressing both LLM lifecycle vulnerabilities and agent-specific architectural risks together. We hope the taxonomies, comparative analyses, and research directions presented here offer a useful foundation for future work toward building reliable, secure, and resilient autonomous AI pentesting systems.

\bibliographystyle{ACM-Reference-Format}
\bibliography{references}

\end{document}